\documentclass{article}

\usepackage{hyperref}
\usepackage[margin=3cm]{geometry}
\usepackage{graphicx}
\usepackage{booktabs}
\usepackage{multirow}
\usepackage{subfig}
\usepackage[round]{natbib}
\usepackage{amsmath}
\usepackage{enumerate}

\newcommand{\degrees}{\ensuremath{^{\circ}}}
\newcommand{\mr}{\multirow}
\newcommand{\mc}{\multicolumn}
\newcommand{\tej}{Tropical Easterly Jet}
\newcommand{\tp}{Tibetan Plateau}
\newcommand{\cam}{CAM-3.1}
\newcommand{\cesm}{CESM-1.0.2}
\newcommand{\cc}{Ctrl}
\newcommand{\cct}{Ctrl$\tau_{d12}$}
\newcommand{\ccz}{CtrlZM}
\newcommand{\cczt}{CtrlZM{$\tau_{d3}$}}
\newcommand{\nglo}{noGlOrog}
\newcommand{\nglot}{noGlOrog$\tau_{d12}$}
\newcommand{\rn}{Reanalysis}
\newcommand{\nc}{NCEP}
\newcommand{\er}{ERA40}
\newcommand{\gp}{GPCP}
\newcommand{\cm}{CMAP}
\newcommand{\pc}{$P_c$}
\newcommand{\zm}{Zhang-MacFarlane}
\newcommand{\um}{$U_{max}$}
\newcommand{\md}{mm day$^{-1}$}
\newcommand{\ms}{m s$^{-1}$}
\newcommand{\td}{$\tau_{d}$}

\newcommand{\de}{\degrees{}E}

\newcommand{\dn}{\degrees{}N}
\newcommand{\ds}{\degrees{}S}

\begin{document}

\title{The Impact of Orography and Latent Heating on the Location of the \tej{}}

\author{S. Rao \\
        \small Department of Mechanical Engineering \\
        \small Indian Institute of Science \\
        \small Bangalore 560012, India \\
        \small \href{mailto:samrat@mecheng.iisc.ernet.in}{samrat@mecheng.iisc.ernet.in} \\
        \small Mobile: +919916675221
\\
        J. Srinivasan \\
        \small Centre for Atmospheric and Oceanic Sciences \\
        \small Indian Institute of Science \\
        \small Bangalore 560012, India
}

\maketitle

\begin{abstract}
The \tej{} (TEJ) is a prominent atmospheric circulation feature observed during the Asian Summer Monsoon. It is generally assumed that Tibet is an essential ingredient in determining the location of the TEJ. However studies have also suggested the importance of latent heating in determining the jet location. The relative importance of Tibetan orography and latent heating is explored through simulations with a general circulation model.

The simulation of TEJ by the Community Atmosphere Model, version 3.1 (\cam{}) has been discussed in detail. Although the simulated TEJ replicated many observed features of the jet, the jet maximum was located too far to the west when compared to observation. The precipitation in the control simulation was high to the west of India and this caused the TEJ to shift westwards by approximately the same amount. Orography was found to have minimal impact on the simulated TEJ hence indicating that latent heating is the crucial parameter. The primacy of latent heating in determining the jet location was confirmed by additional simulations where the simulated precipitation was brought closer to observations. This made the TEJ to also shift to the correct position.
\end{abstract}

\section{Introduction}\label{intro}

The \tej{} (TEJ) is one of the most defining features of the Indian Summer Monsoon (ISM) which itself is a part of the Asian summer monsoon (ASM). The jet is observed mostly during the ISM that is in the months of June to September. It has a maximum between 50\de{}-80\de{}, Equator-15\dn{} and at 150hPa. The TEJ was first discovered by \citet{koteswaram-58}. It has a great influence on the rainfall in Africa (\citealp{hulme-89}). The correct simulation of the TEJ is important for accurate seasonal predictions and weather forecasting.

The \tej{} is believed to be influenced by the \tp{}. Previous studies (\citealp{flohn-68}, \citealp{krishnamurti-71}) have highlighted the presence of a huge upper tropospheric anticyclone above the \tp{} in summer. The origin of the Tibetan anticyclone itself has been attributed to the summertime insolation on the \tp{}. This sensible heating is widely regarded as the primary reason for this region to act as an elevated heat source. \cite{flohn-65} first suggested that in summer, southern and southeastern Tibet, i.e. south of 34\dn{}-35\dn{}, act as an elevated heat source which changes the meridional temperature and pressure gradients and contributes significantly to the reversal of high tropospheric flow during early June. \cite{flohn-68} showed that sensible heat source over the \tp{} as well as latent heat release due to monsoonal rains over central and eastern Himalayas and their southern approaches generates a warm core anticyclone in the upper troposphere at around 30\dn{}. This was a primary mechanism for establishing the south Asian monsoonal circulation over south Asia. According to Koteswaram these winds are a part of the Tibetan Anticyclone which forms during the summer monsoon over South Asia. He believed that the southern flank of the anticyclone preserved its angular momentum and became an easterly at approximately 15\dn{}. Contradicting Koteswaram, \cite{raghavan-73} opined that the upper tropospheric zonal component of the equatorward outflow from the Tibetan anticyclone does not agree with the law of conservation of angular momentum. However the thermal gradient balance was applicable and this was responsible for the origin of the TEJ.

The TEJ is not simply a passive atmospheric phenomena. \cite{hulme-89} studied the impact of the TEJ on Sudan rainfall. El Ni\~{n}o events were suggested as a direct control over Sahelian rainfall via the TEJ. The decelerating limb of the TEJ showed interannual variations in location in relation to Eastern Pacific warming. Fluctuations in the jet were shown to be responsible for altering precipitation patterns in Sudan. \cite{camberlin-95} also reported on the significant linkages between interannual variations of summer rainfall in Ethiopia-Sudan region and strength and latitudinal extent of upper-tropospheric easterlies. More recently \cite{nicholson-07} showed that the wave activity of the TEJ influences the African easterly jet (AEJ). The AEJ in turn influences the weather patterns over the Atlantic coast of Africa. \cite{bordoni-08} discussed the role of TEJ in modifying the monsoonal circulation. Upper-level easterlies shield the lower-level cross-equatorial monsoonal flow (which later becomes the Somali jet) from extratropical eddies. This made the angular momentum conservation principle applicable to overturning Hadley cell dynamics. In the larger picture this implies a strengthening of the monsoonal regime. This shows that the TEJ actively shapes the climate and weather pattern in regions not under the direct influence of the Indian Summer Monsoon.

Thus we can see that while there have been attempts in the past to explain some aspects of the TEJ and its effects, there has been no systematic study to understand the role of orography and different heat sources that set up the thermal gradients that form and maintain the TEJ both in strength and spatial location. We have used an AGCM to study the impact of orography and monsoonal heat sources on the TEJ. The jet location and its response to have been studied by:
\begin{enumerate}[(i)]
\item Modifying the deep-convective scheme and thereby changing the monsoonal heating pattern.
\item Removing orography.
\end{enumerate}

\section{The \tej{} in \rn{}}\label{tej-rean}

\begin{figure}[htbp]
   \begin{center}
   \includegraphics[trim = 5mm 22mm 55mm 70mm, clip, scale=0.5]{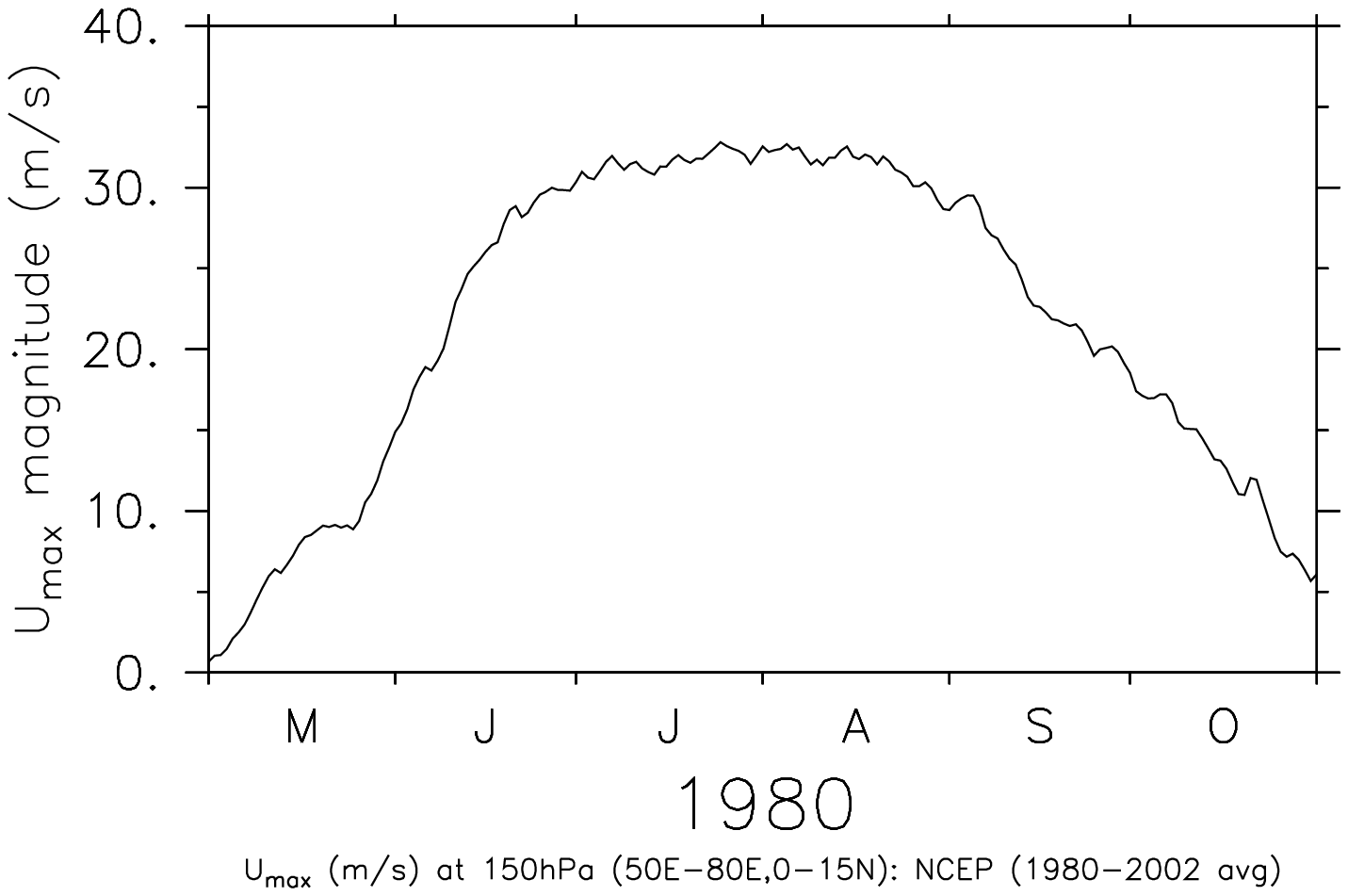}
   \end{center}
   \caption{Time series of 150 hPa zonal wind averaged between 50\de{}-80\de{}, Equator-15\dn{} showing TEJ peaking in July-August (\nc{}: 1980-2002 avg)}
   \label{fig: nc-U150xyavg-t}
\end{figure}

We use the standard \rn{} \nc{} (\citealp{kalnay-96}), \er{} (\citealp{uppala-05}) data for the winds and \gp{} (\citealp{adler-03}), \cm{} (\citealp{xie-97}) data for precipitation. Unless otherwise mentioned the time period chosen spans years 1980 to 2002. The maximum zonal winds occur at 150hPa. From Fig. \ref{fig: nc-U150xyavg-t} we see that the TEJ peaks in July to August. In the present work we have focused on the TEJ during the month of July when it first reaches its maximum value.

\begin{figure}[htbp]
   \begin{center}
   \subfloat[Horizontal section (150 hPa)]{\label{nc-Umax-xy}\includegraphics[trim = 5mm 20mm 10mm 60mm, clip, scale=0.35]{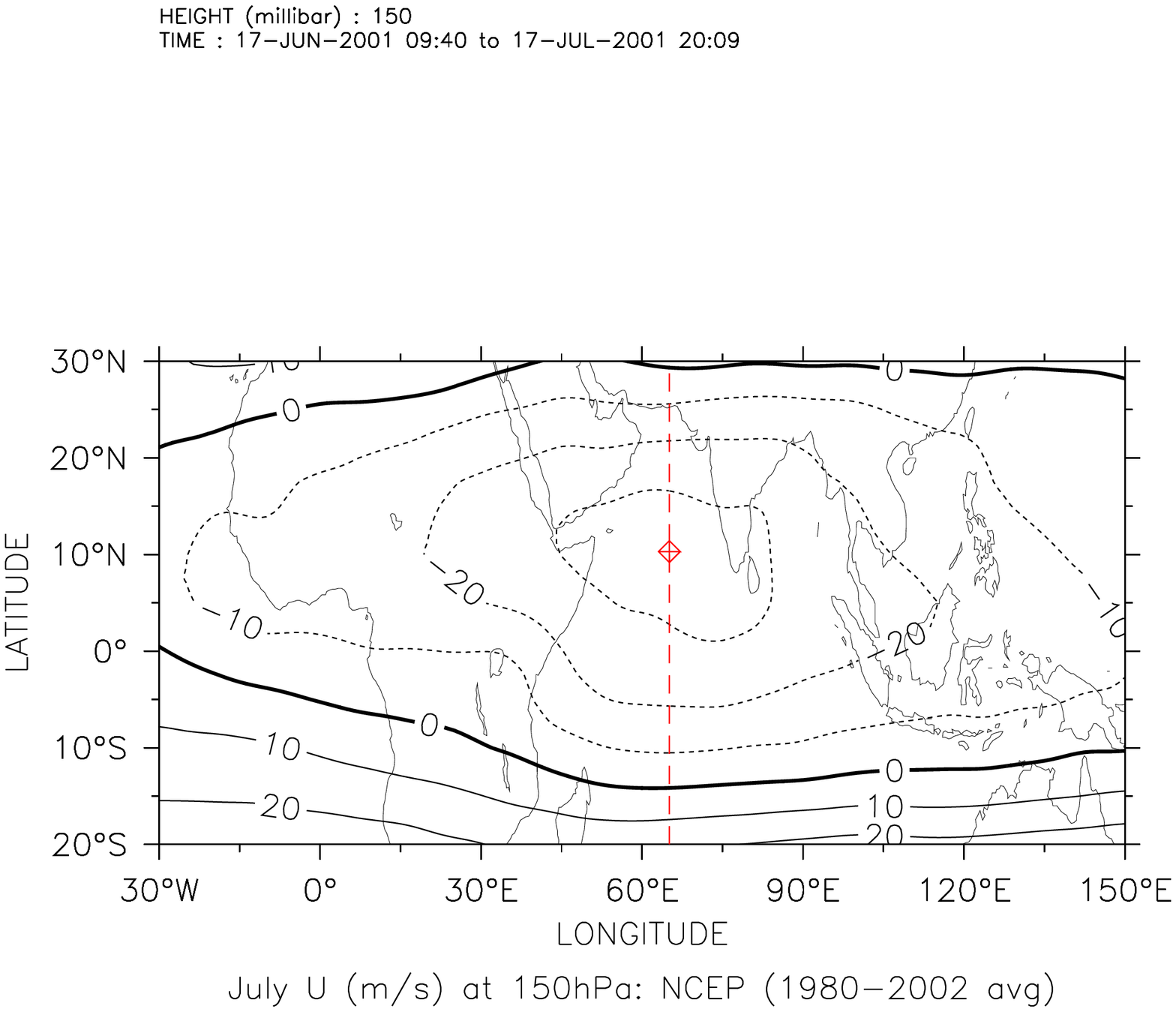}}
   \hspace{5mm}
   \subfloat[Meridional section (65.1\de{})]{\label{nc-yz}\includegraphics[trim = 0mm 15mm 80mm 65mm, clip, scale=0.35]{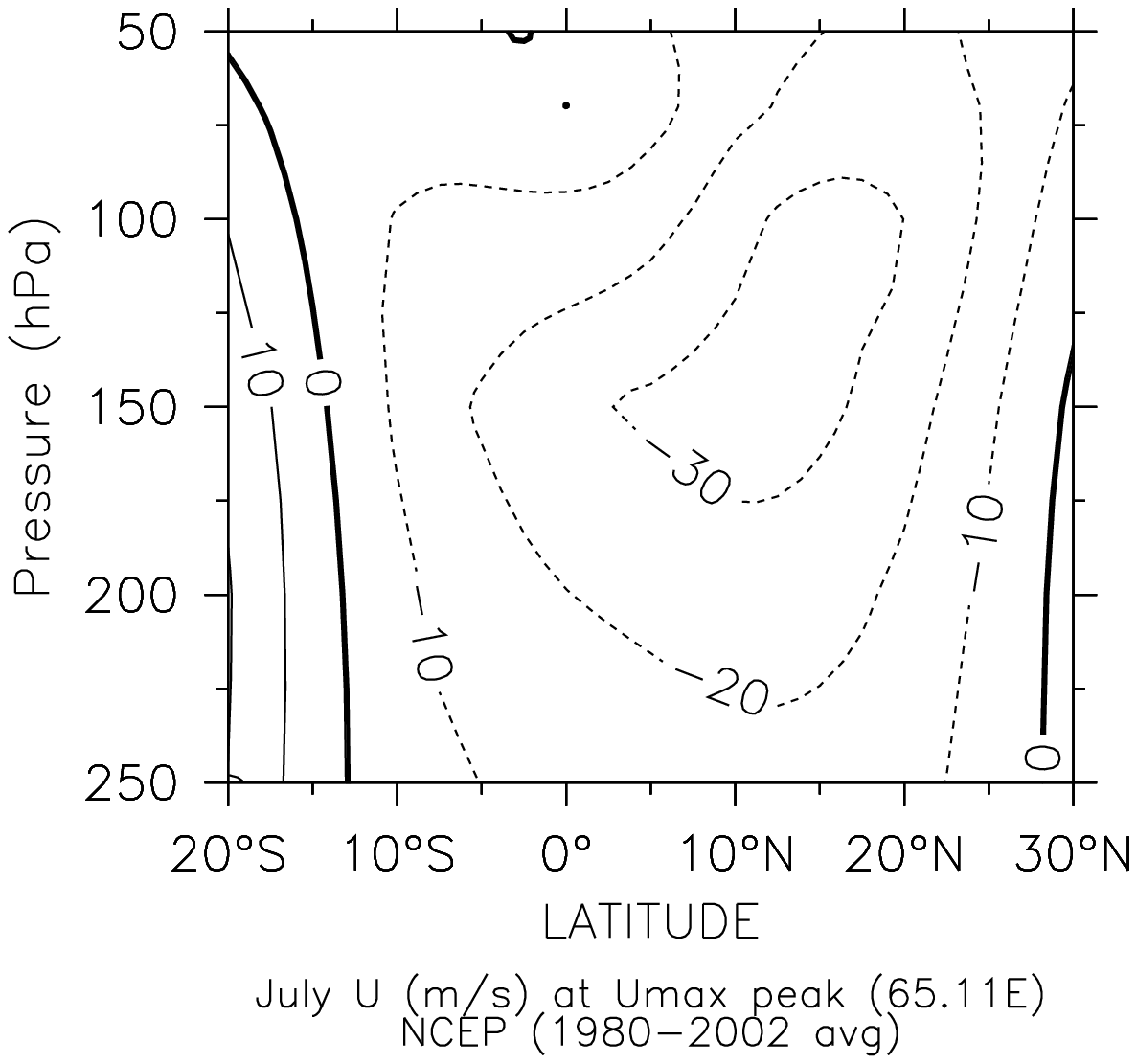}}
   \end{center}
   \caption{\protect\subref{nc-Umax-xy} July zonal wind at pressure level where \um{} is attained, `cross-diamond' is location of peak zonal wind; \protect\subref{nc-yz} July zonal wind at longitude where \um{} is attained. \nc{} (1980-2002 avg).}
   \label{fig: nc-jul-Umax-xy}
\end{figure}

We have listed the magnitude of peak zonal wind and its location in table \ref{tab: mix-Umax,R}. Due to the good agreement between zonal winds of \nc{} and \er{} we have only shown the former. We show, in Fig. \ref{nc-Umax-xy}, the zonal wind and location of peak zonal wind (\um{}) for \nc{} at 150hPa. The `cross-diamond' indicates the horizontal location of \um{}. The location of peak zonal wind of each individual year from 1980-2002 was found and then the means were calculated. The standard deviation for the zonal winds for \nc{} and \er{} was found to be 2\ms{} and 1.7\ms{} respectively. The meridional section of the zonal wind is shown in Fig. \ref{nc-yz}. This vertical cross-section corresponds to the longitude where the zonal wind is maximum. The meridional structure shows the jet peak lying between equator and $\sim$20\dn{}. We also observe a vertical equator to pole tilt. The mean height of maximum easterly zonal wind speeds are higher at higher latitudes. In Fig. \ref{nc-z3-xy} the velocity vectors and geopotential high is shown. The peak is not over the \tp{} but to the west of it.

\begin{table}[htbp]
\caption{Magnitude (\ms{}) and location of peak zonal wind, and centroid of mean precipitation}
\label{tab: mix-Umax,R}
\begin{center}
\begin{tabular}{ccccccc} \toprule
\mc{7}{c}{\textbf{\rn{} and observation}} \\
\mr{2}{*}{Case} & \mc{4}{c}{Zonal wind}                 & \mc{2}{c}{Precipitation} \\ \cline{2-7}
                & Peak  & Lon       & Lat       & Press & Lon       & Lat \\ \midrule
\nc{}           & 34.48 & 65.1\de{} & 10.3\dn{} & 150 & \\
\er{}           & 33.61 & 68.7\de{} & 10.4\dn{} & 150 & \\
\gp{}           &       &           &           &       & 96.0\de{} & 11.9\dn{} \\
\cm{}           &       &           &           &       & 95.8\de{} & 10.0\dn{} \\ \midrule \midrule
\mc{7}{c}{\textbf{\cam{} simulations}} \\
\mr{2}{*}{Case} & \mc{4}{c}{Zonal wind}                 & \mc{2}{c}{Precipitation} \\ \cline{2-7}
                & Peak  & Lon       & Lat       & Press & Lon       & Lat \\ \midrule
\cc{}           & 47.19 & 42.5\de{} & 8.8\dn{}  & 125   & 76.7\de{} & 9.7\dn{} \\
\nglo{}         & 43.45 & 43.5\de{} & 7.6\dn{}  & 125   & 77.0\de{} & 7.4\dn{} \\
\cct{}          & 48.79 & 64.0\de{} & 7.2\dn{}  & 125   & 93.4\de{} & 14.8\dn{} \\
\nglot{}        & 39.51 & 64.0\de{} & 6.0\dn{}  & 125   & 92.5\de{} & 12.1\dn{} \\
\ccz{}          & 43.12 & 49.5\de{} & 5.6\dn{}  & 130   & 77.6\de{} & 12.1\dn{} \\
\cczt{}         & 46.28 & 60.5\de{} & 6.4\dn{}  & 125   & 93.6\de{} & 14.8\dn{} \\ \midrule
\mc{7}{c}{\cc{}, \cct{}, \ccz{}, \cczt{}: default orography} \\
\mc{7}{c}{\nglo{}, \nglot{}: no orography} \\ \bottomrule
\end{tabular}
\end{center}
\end{table}

\subsection{Impact of latent heating on the location of the TEJ}\label{tej-latent}

The TEJ is not a stationary entity. Variations in its position in different years can also be observed. This is most strikingly observed in July 1988 and 2002. The location of the TEJ in during these two years is radically different. This is seen in Fig. \ref{nc-jul-Umax-88,02-xy} where the 30\ms{} zonal wind contour and \um{} locations show significant differences. In fact the location of \um{} is shifted eastwards by $\sim$20\degrees{} in 2002. These differences are seen in \er{} data as well. The vertical structure in these two years is similar to Fig. \ref{nc-yz}.

\begin{figure}[htbp]
   \begin{center}
   \subfloat[\nc{} zonal wind, 1988 (red), 2002 (blue)]{\label{nc-jul-Umax-88,02-xy}\includegraphics[trim = 5mm 20mm 10mm 60mm, clip, scale=0.35]{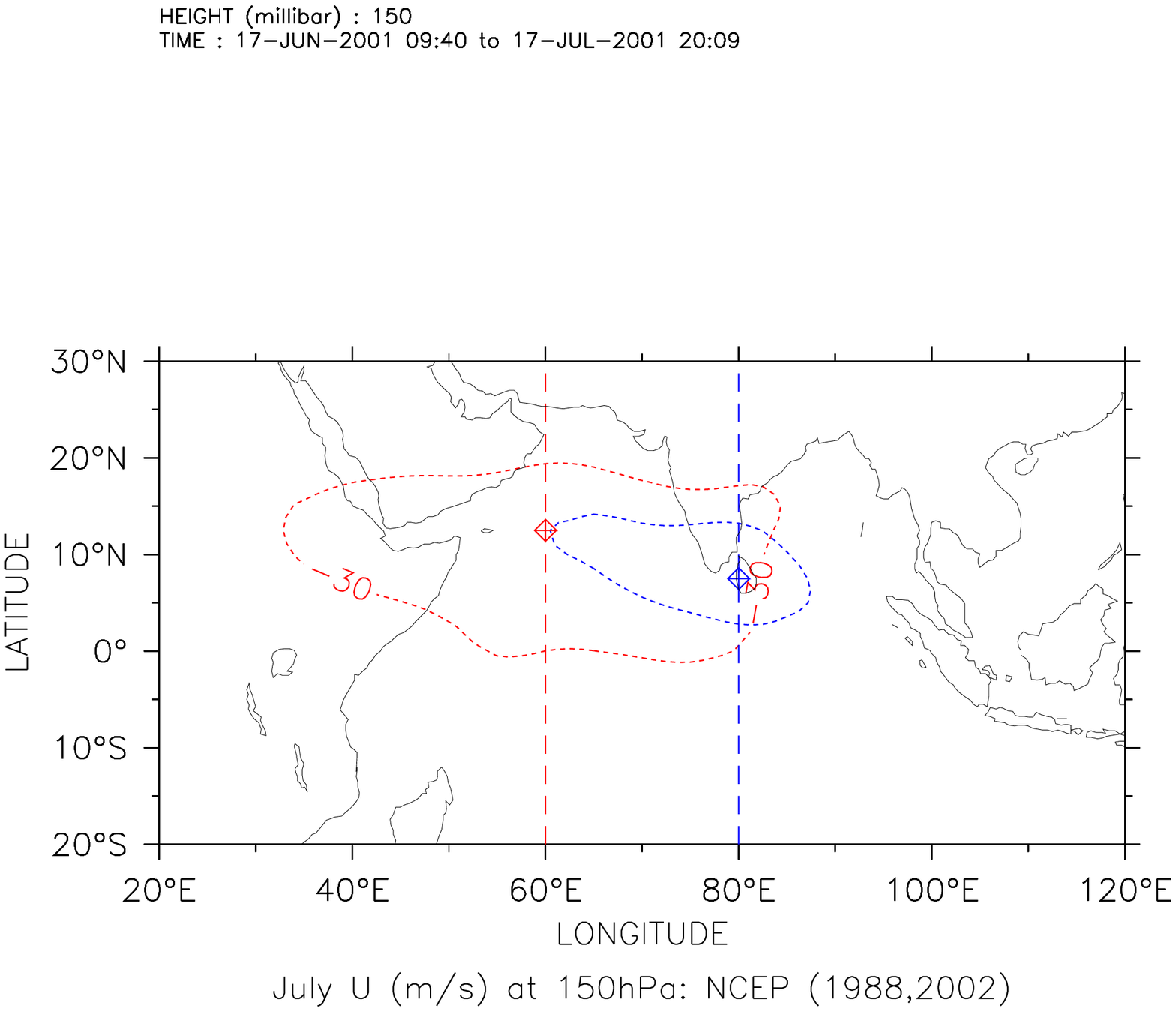}}
   \hspace{5mm}
   \subfloat[\nc{} precipitation difference: 1988--2002]{\label{nc-jul-Rdiff-xy}\includegraphics[trim = 5mm 20mm 0mm 60mm, clip, scale=0.35]{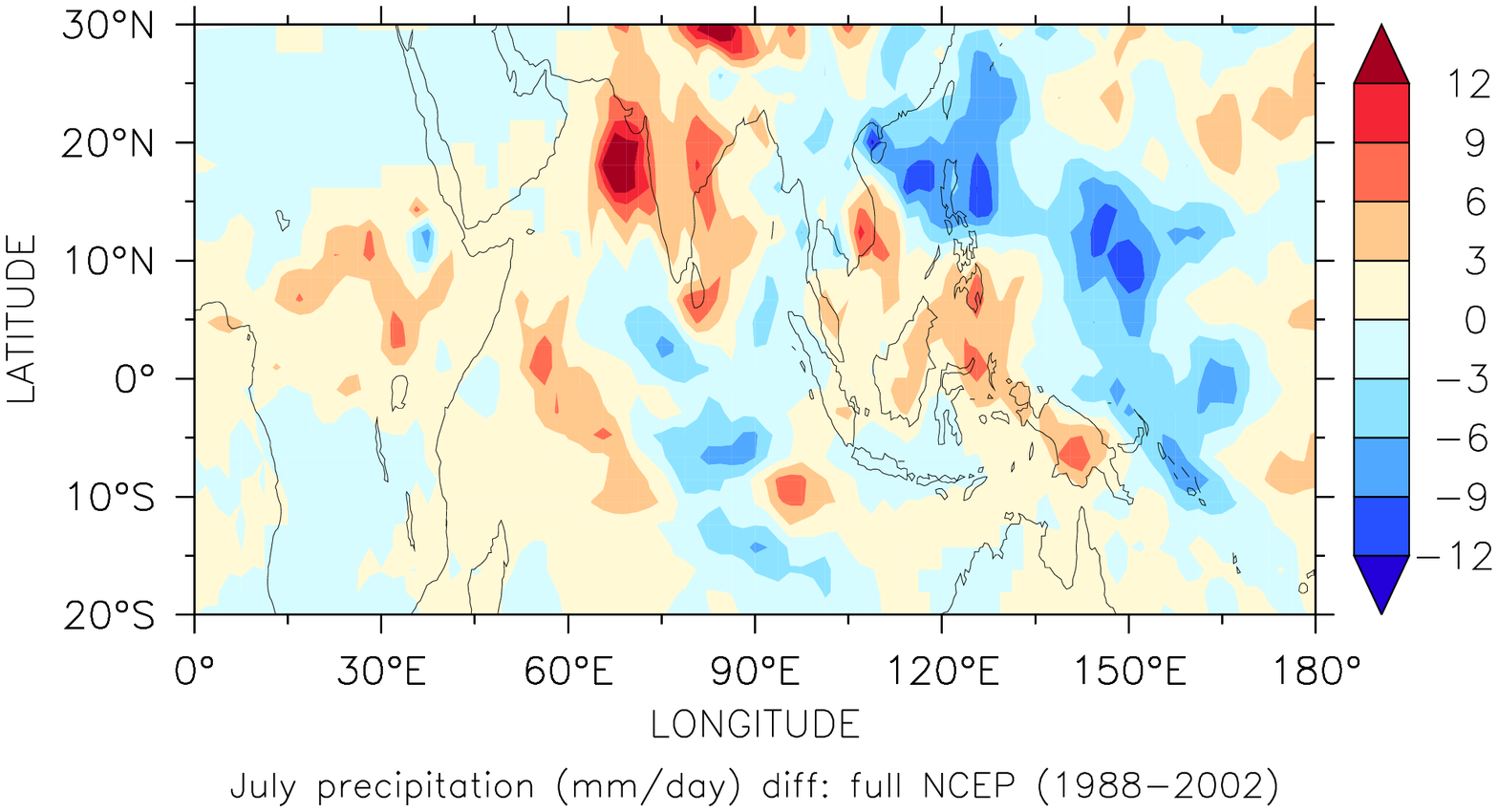}}
   \end{center}
   \caption[July precipitation and zonal wind: \nc{} in 1988 and 2002]{\protect\subref{nc-jul-Umax-88,02-xy} 30\ms{} zonal wind contour (dashed) in 1988 and 2002 showing shift in TEJ in response to heating, cross-sections are at \um{} pressure level (150hPa), `cross-diamond' shows the location of maximum zonal wind; \protect\subref{nc-jul-Rdiff-xy} precipitation difference (\md{}), year 2002 subtracted from 1988. All data from \nc{}.}
   \label{fig: nc-R,Umax-xy}
\end{figure}

We next highlight the importance of latent heating on the location of the jet. It is well known that in India 2002 was a drought year (\citealp{sikka-03}, \citealp{bhat-06}) while 1988 was an excess monsoon year. The relationship between precipitation and the jet location in the month of July for these two years is now clarified. Here in order to be self-consistent we have used both precipitation and zonal winds from \nc{} data. This is because we expect that \rn{} data is self-consistent and that the dynamic response of the atmosphere is entirely on account of forcing from the same data. If the TEJ is influenced largely by latent heating due to Indian Summer Monsoon then the July 2002 shift should be due to the mean rainfall being eastward shifted. This is clearly seen in Fig. \ref{nc-jul-Rdiff-xy} where we have shown precipitation differences between July 1988 and 2002. The differences are quite striking. In July 2002 there is a clear eastward shift in rainfall with the maximum being in the Pacific warm pool and relatively little in the Indian region. This is in contrast with July 1988 where Indian region received high amounts of precipitation. The jet in 1988 peaks at 60\de{}-65\de{} while in 2002 the maximum is in the southern Indian peninsula implying a $\sim$20\degrees{} westward shift. We have also verified the same in \er{} data and found that the behavior was the same for these two years.

In fact the TEJ also shows intraseasonal and interdecadal variations in its location. \cite{sathiyamoorthy-07} reported that the axis of the jet undergoes meridional movement in response to active and break phases of the Indian Monsoon. \cite{sathiyamoorthy-05} also observed a reduction in the spatial extent of the TEJ between 1960-1990. According to him the TEJ almost disappeared over the Atlantic and African regions. This reduction coincided with the 4-5 decade prolonged drought conditions over the Sahel region. He speculated that these two phenomena were associated with each other. This explanation also implies that latent heating significantly influences the characteristics of the TEJ.

We also mention here that \cite{zhang-02} had found that the South Asian High had a bimodal structure. Its two modes -- one the Tibetan mode and the other the Iranian mode, both of which were fairly regular in their occurrence, were mostly influenced by heating effects. The former owed its existence to diabatic heating of the \tp{}, while the latter occurred due to adiabatic heating in the free atmosphere and diabatic heating near the surface.

However the issue about the importance of orography on the TEJ has not yet been settled. In the previous discussions variations in the jet were in the presence of the Himalayas and \tp{}. Although the importance of latent heating on the TEJ is quite clear now we need to confirm the importance of orography in deciding the location of the TEJ. The importance of orography especially \tp{} in setting up the thermal gradients necessary for the formation of the TEJ has been discussed by many authors. \cite{ye-81} did laboratory experiments to simulate the heating effect of elevated land. He introduced heating in an ellipsoidal block resulting in vertical circulation, an anticyclone in the upper layer and a cyclonic flow in the lower layer. The pattern was found to be qualitatively similar to summer time atmospheric circulation in south Asia. \cite{raghavan-73} discussed the importance of \tp{} and explained that the jet owed its existence the the temperature gradients that was partly influenced by the Tibetan high. According to \cite{wang-06} the elevated heat source of the \tp{} was instrumental in providing an anchor to locate the Tibetan high. On the other hand \cite{hoskins-95} and \cite{liu-07} have argued that orography plays a secondary role in determining the position of the summertime upper tropospheric anticyclone. \cite{jingxi-89} studied the TEJ at 200 hPa and found that precipitation changes in the west coast of India led to changes in the jet structure. \cite{chen-87} observed that at 200 hPa the jet was weaker during El Ni\~{n}o and Indian drought events. The TEJ was weaker in the drought years of 1979, 1983 and 1987 and stronger in the excess monsoon years 1985 and 1988.

In light of these considerable differences of opinion it is necessary to study the jet in a GCM and understand the relative roles played by latent heating and orography on the TEJ. We discuss this approach below.

\section{The \tej{} in \cam{}}\label{tej-cam}

\subsection{Description of \cam{} and experiment details}\label{cam-exp}

The AGCM that has been used for the present work is the Community Atmosphere Model, version 3.1 (\cam{}). The finite-volume dynamical core using the recommended 2\degrees{}$\times$2.5\degrees{} grid resolution has been used for all simulations. The time step is 30 minutes and 26 levels in the vertical are used. Deep convection is the \cite{zhang-95} scheme while shallow convection is the \cite{hack-94} scheme. Stratiform processes employs the \cite{rasch-98} scheme updated by \cite{zhang-03}. Cloud fraction is computed using a generalization of the scheme introduced by \cite{slingo-89}. The shortwave radiation scheme employed is described in \cite{briegleb-92}. The longwave radiation scheme is from \cite{ramanathan-86}. Land surface fluxes of momentum, sensible heat, and latent heat are calculated from Monin-Obukhov similarity theory applied to the surface. Climatological mean SST was specified as the boundary condition. Sea surface temperatures are the blended products that combine the global Hadley Centre Sea Ice and Sea Surface Temperature (HadISST) dataset (\citealp{rayner-03}) for years up to 1981 and \cite{reynolds-02} dataset after 1981.

We ran the model in its default configuration for a five year period. This simulation is referred to as the control (\cc{}) simulation. Additionally another simulation (referred to as \nglo{}) has been conducted to check the influence of orography on the TEJ. This has also been run for five years with same boundary conditions but with orography all over the globe removed. This latter simulation is used to investigate the direct influence of topography on the TEJ. All the simulation results presented in this paper are based on five year means.

\subsection{Impact of orography on the simulated TEJ}\label{cam-orog}

\begin{figure}[htbp]
   \begin{center}
   \subfloat[\cc{} (125 hPa)]{\label{cc-Umax-xy}\includegraphics[trim = 5mm 20mm 10mm 60mm, clip, scale=0.35]{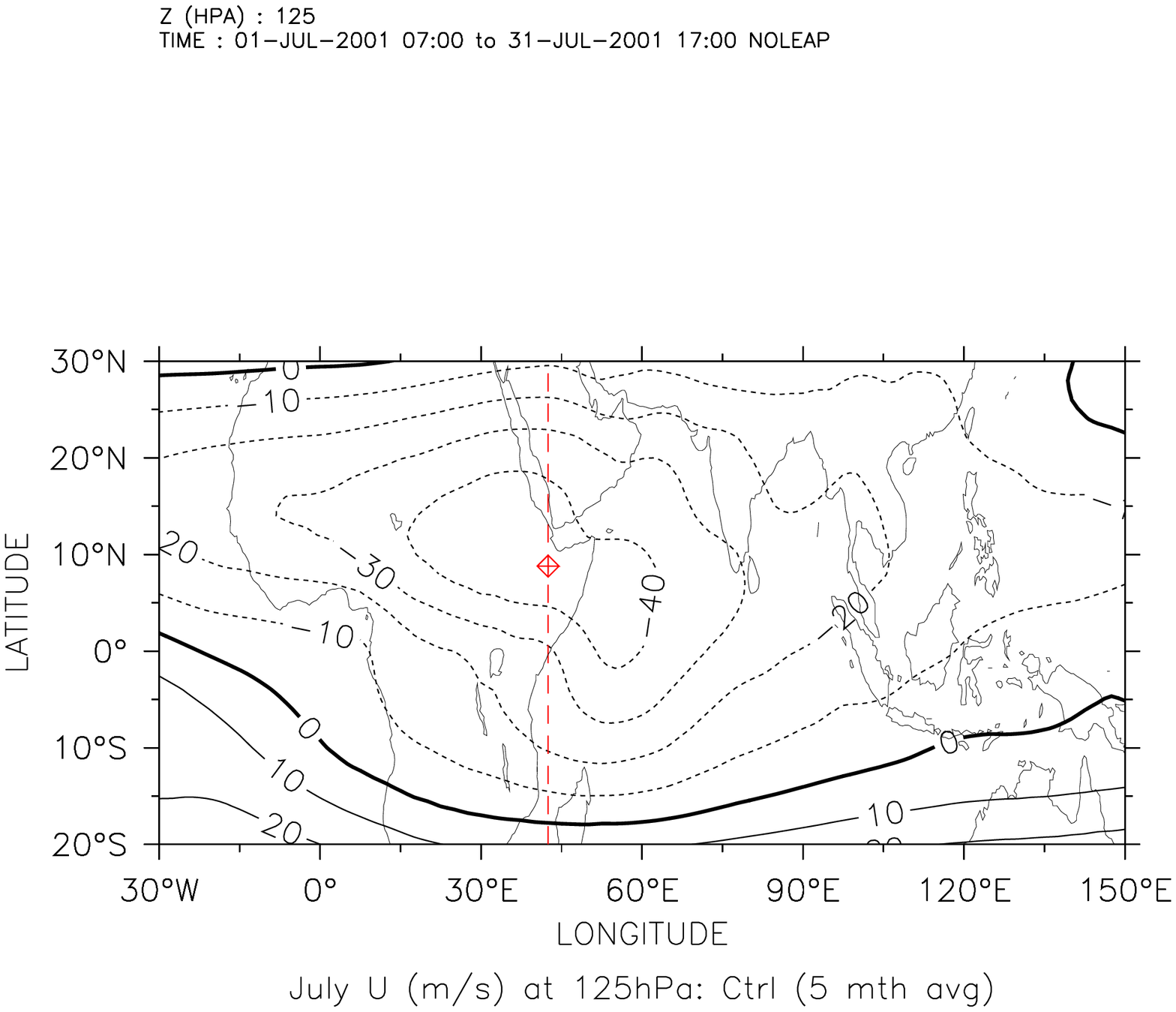}}
   \hspace{5mm}
   \subfloat[\cc{} (42.5\de{})]{\label{cc-yz}\includegraphics[trim = 0mm 15mm 80mm 65mm, clip, scale=0.35]{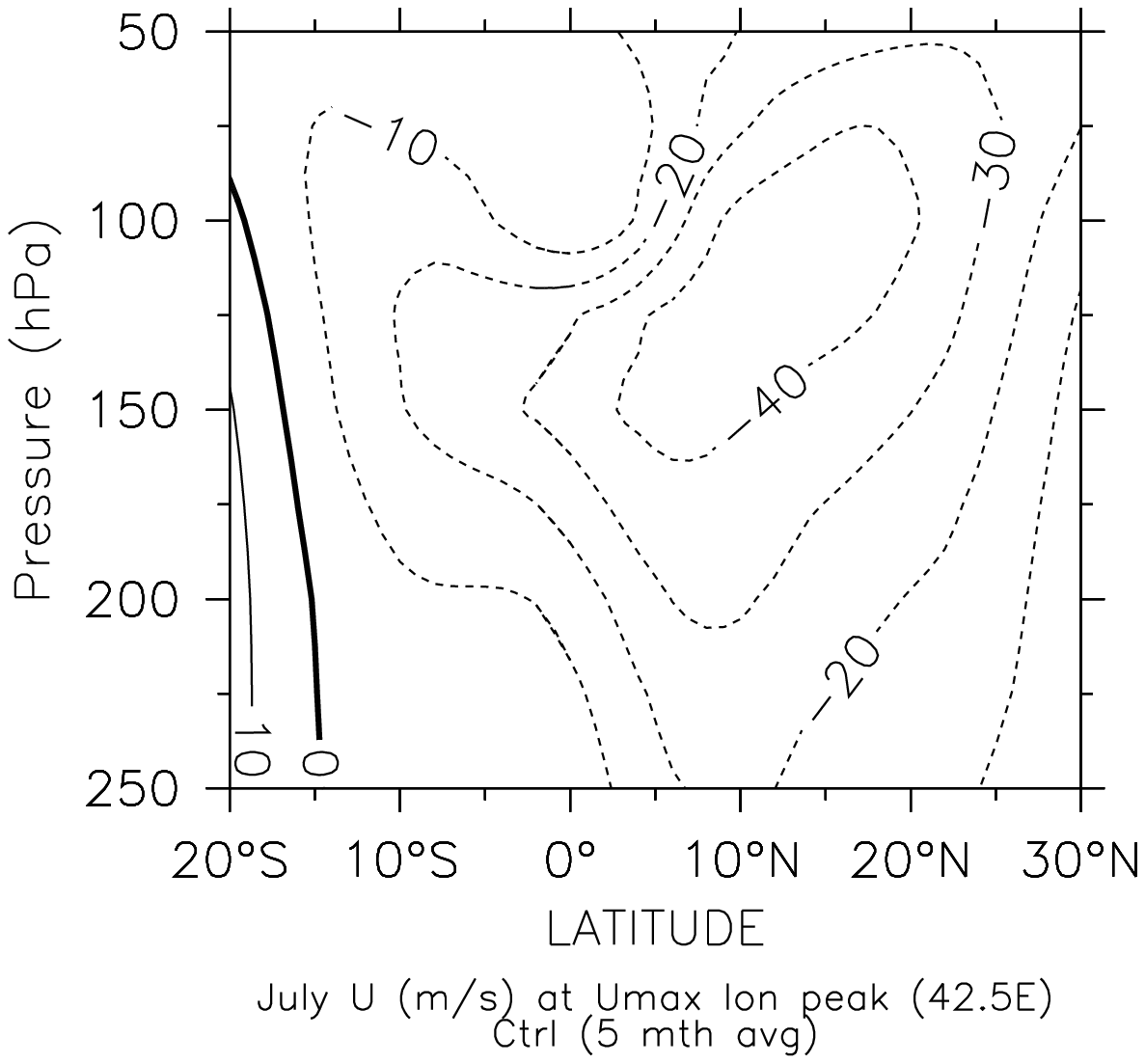}}
   \vskip 5mm
   \subfloat[\nglo{} (125 hPa)]{\label{nglo-Umax-xy}\includegraphics[trim = 5mm 20mm 10mm 60mm, clip, scale=0.35]{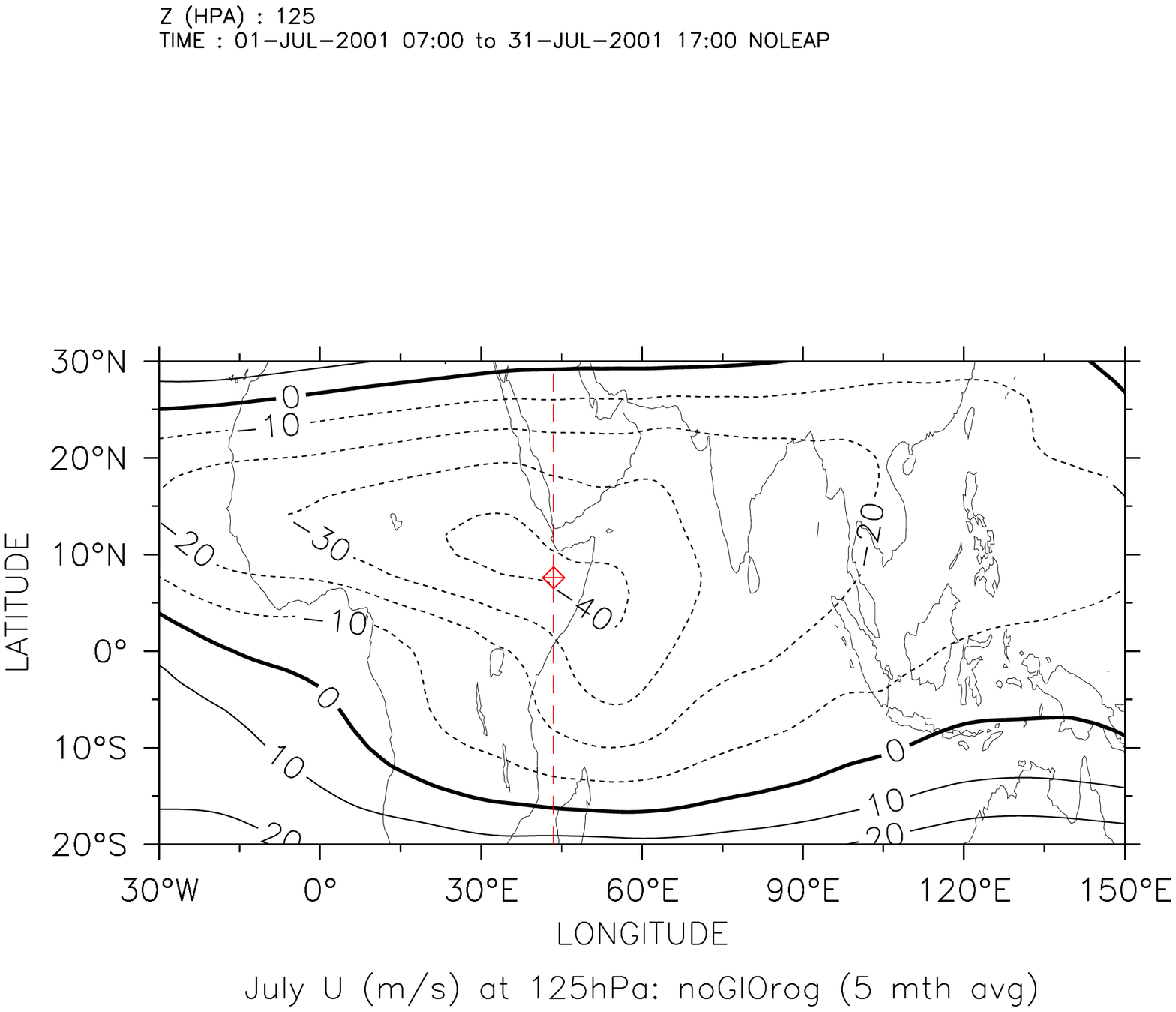}}
   \hspace{5mm}
   \subfloat[\nglo{} (43.5\de{})]{\label{nglo-yz}\includegraphics[trim = 0mm 15mm 80mm 65mm, clip, scale=0.35]{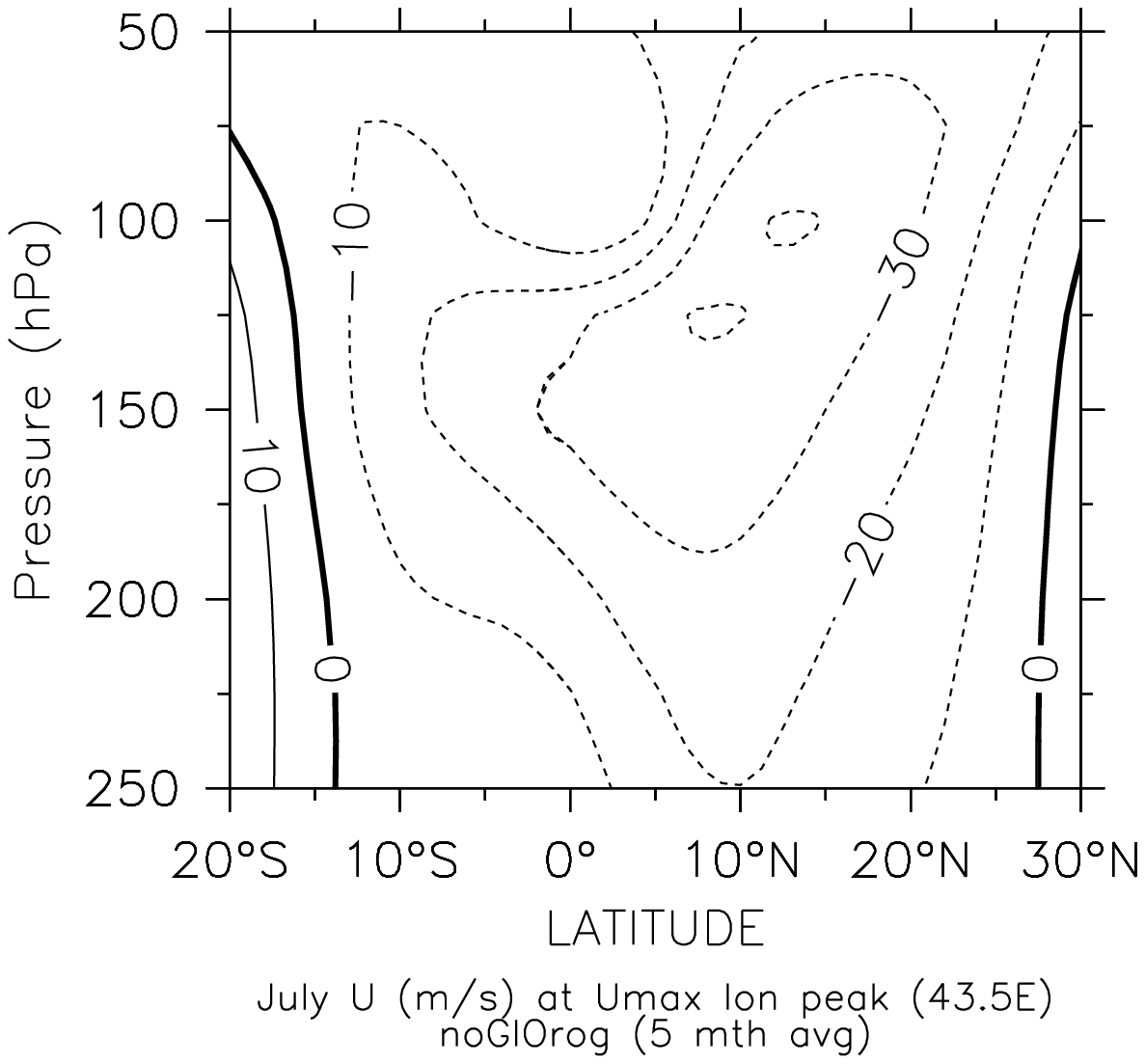}}
   \end{center}
   \caption[July meridional and horizontal zonal wind profile at \um{} location: \cc{} and \nglo{}]{\protect\subref{cc-Umax-xy}, \protect\subref{nglo-Umax-xy} July zonal wind at pressure level where \um{} is attained, `cross-diamond' is location of peak zonal wind; \protect\subref{cc-yz}, \protect\subref{nglo-yz} July zonal wind at longitude where \um{} is attained (refer table \ref{tab: mix-Umax,R}).}
   \label{fig: cam-jul-Umax-xy}
\end{figure}

We have computed the 5 year average of maximum zonal wind and its corresponding location. We show the horizontal (Figs. \ref{cc-Umax-xy},\subref*{nglo-Umax-xy}) and meridional (Figs. \ref{cc-yz},\subref*{nglo-yz}) profiles of the TEJ for \cc{} and \nglo{} simulations. As with \rn{}, the vertical cross-section is at the location of zonal wind maximum. We observe the existence of a \tej{}. In both the simulations the first noticeable feature is a $\sim$30\degrees{} westward shift of the simulated jet. This shift in the default \cc{} was also documented by \cite{hurrell-06} where they analyzed the 200hPa JJA zonal wind fields. The zonal extent of the simulated TEJ is also much higher than \rn{}. The meridional cross-section is very similar to \rn{} and shows a vertical equator to pole tilt with the maximum lying between equator and $\sim$20\dn{}. The depth is also maximum on the poleward side.

The impact of orography in determining the location and spatial structure of the jet in \cam{} is thus minimal. The location of the peak zonal wind is virtually the same for \cc{} and \nglo{} simulations while the jet is weaker in the absence of orography. Although CAM-3.1 does show reasonable fidelity in determining the spatial features of the TEJ, the discrepancies in location and magnitude of the maximum velocity of the TEJ need to be understood. With the insight gained from section \ref{tej-latent} we look at the spatial distribution of simulated and observed rainfall.

\section{Spatial distribution of heating in \rn{} and \cc{}}\label{precip}

The role of Tibetan orography was studied by \cite{boos-10} where they showed that only the narrow orography of the Himalayas and adjacent mountain ranges, and not Tibet, influenced the mean 850hPa winds. However the upper tropospheric winds were not studied in their work. The westward shift of the simulated TEJ indicates that sensible heating from the \tp{} may not play a major role in the existence and location of TEJ. During the monsoon season the major source of heating in the tropics is latent heating and hence we need to look at the role of latent heating.

\begin{figure}[htbp]
   \begin{center}
   \subfloat[Precipitation: \gp{}]{\label{gpcp-jul-R-xy}\includegraphics[trim = 5mm 20mm 0mm 60mm, clip, scale=0.35]{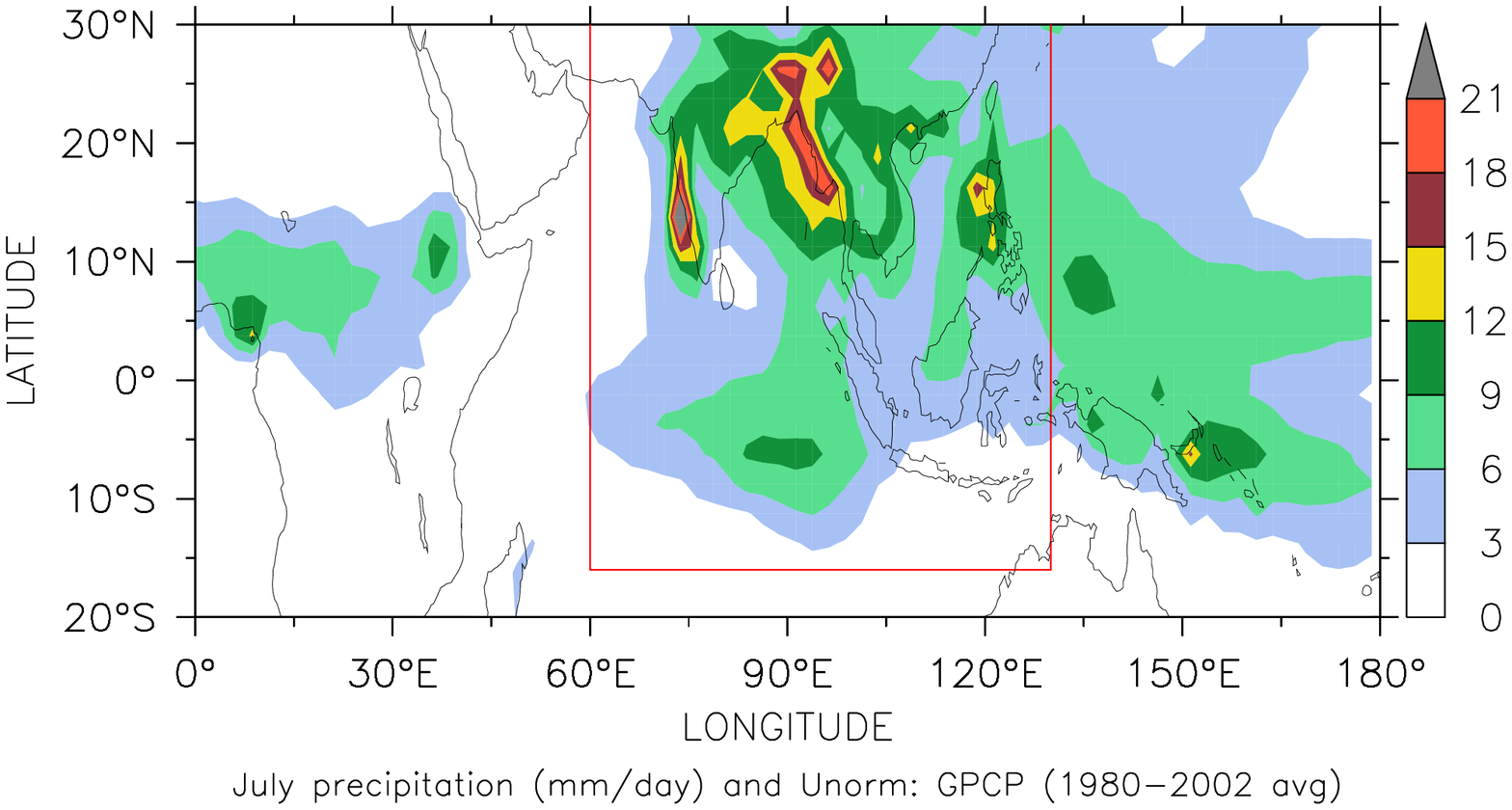}}
   \hspace{5mm}
   \subfloat[Precipitation: \cm{}]{\label{cmap-jul-R-xy}\includegraphics[trim = 5mm 20mm 0mm 60mm, clip, scale=0.35]{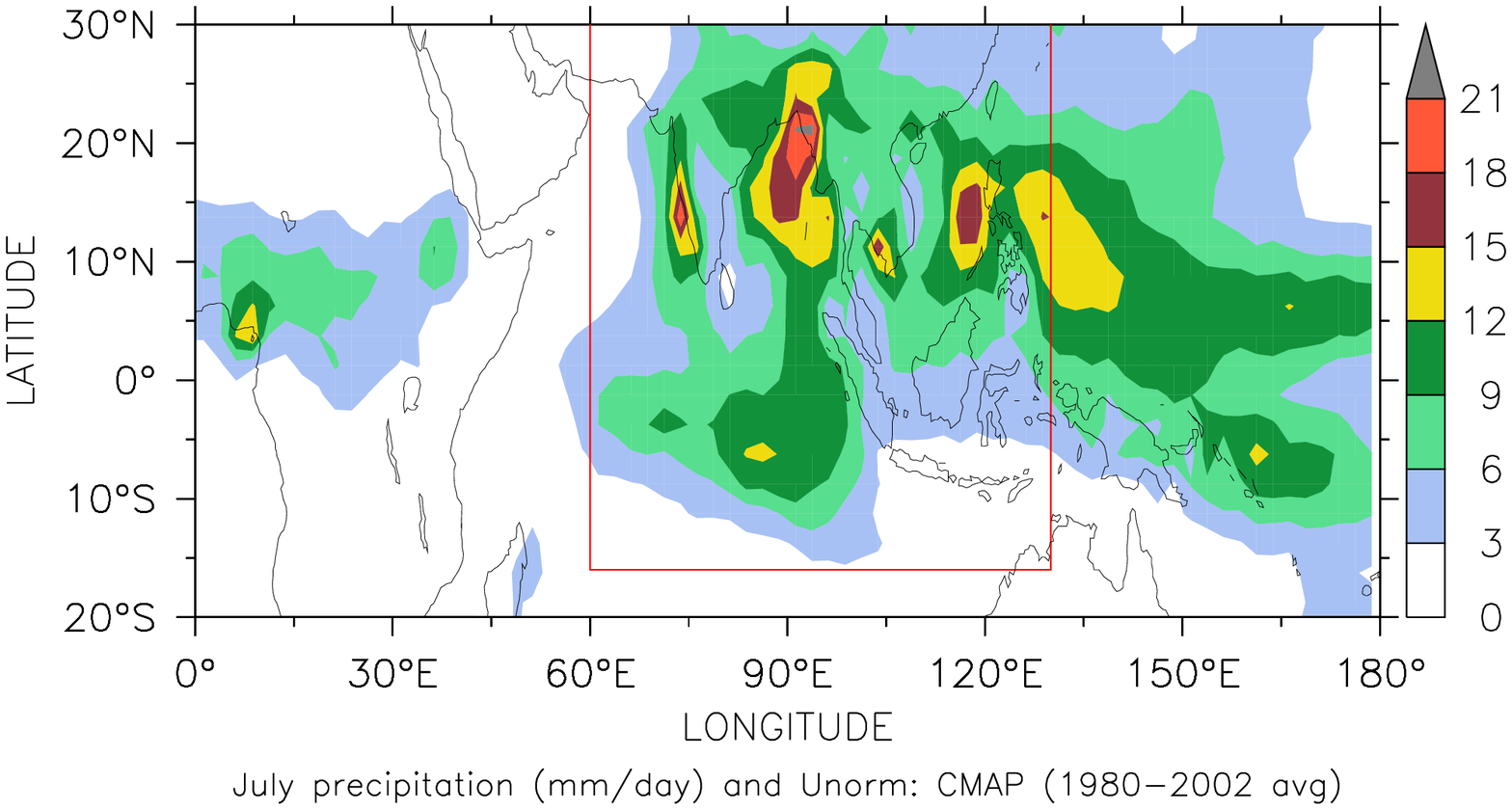}}
   \vskip 5mm
   \subfloat[Precipitation: \cc{}]{\label{cc-jul-R-xy}\includegraphics[trim = 5mm 20mm 0mm 60mm, clip, scale=0.35]{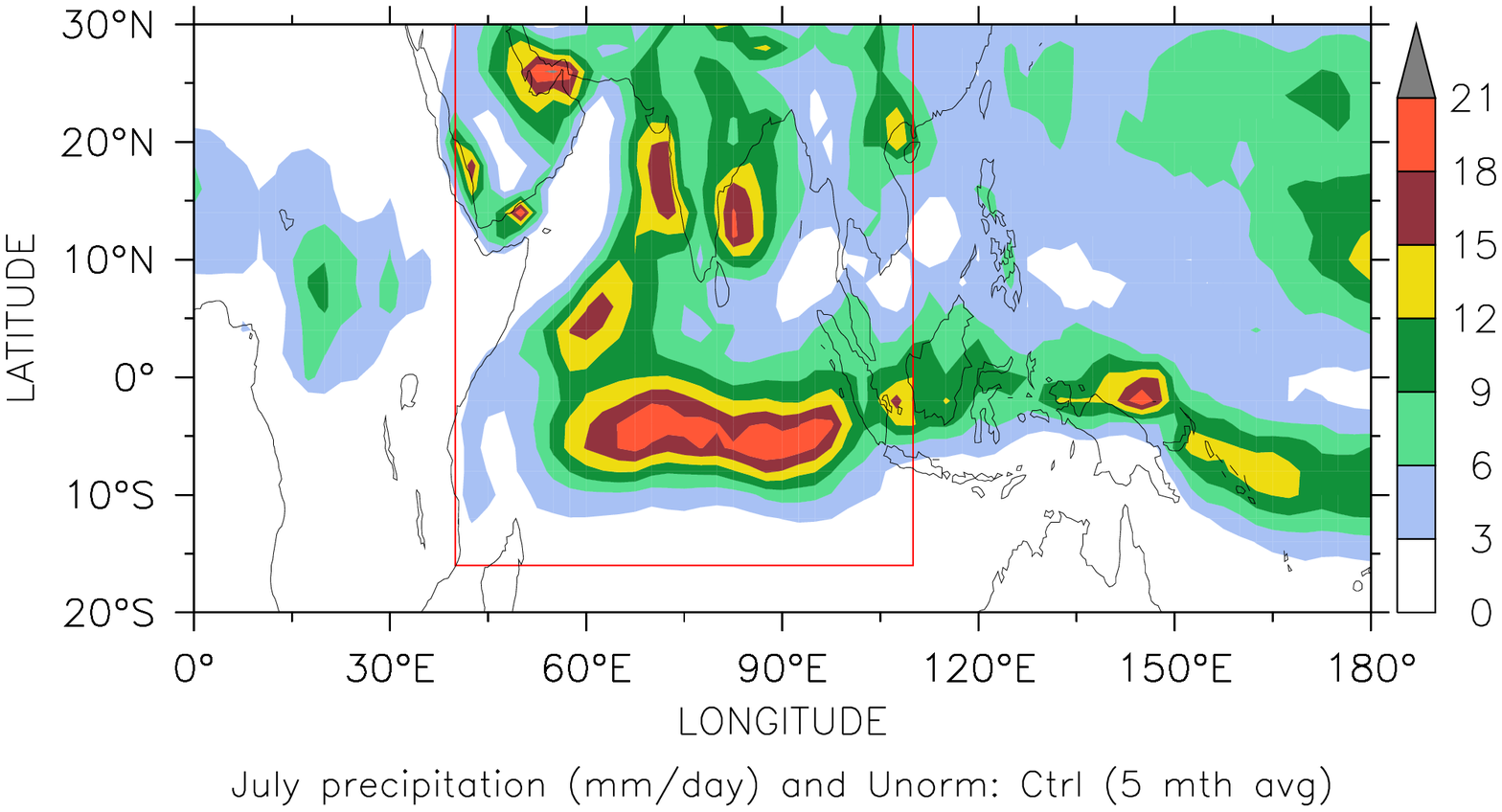}}
   \hspace{5mm}
   \subfloat[Precipitation difference]{\label{mix-Rdiff-xy}\includegraphics[trim = 5mm 20mm 0mm 60mm, clip, scale=0.35]{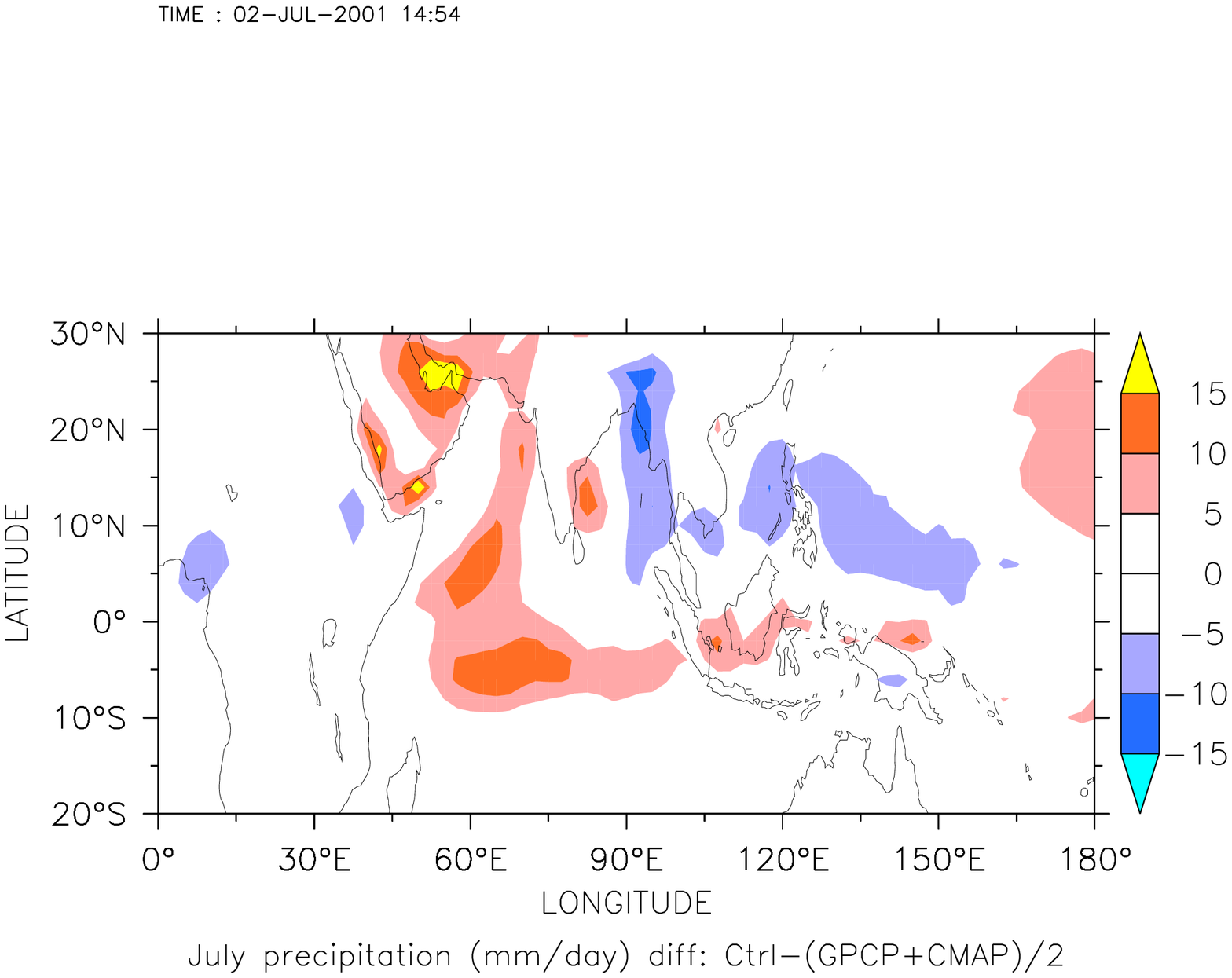}}
   \end{center}
   \caption[July precipitation: \gp{}, \cm{}, \cc{}]{\protect\subref{gpcp-jul-R-xy}, \protect\subref{cmap-jul-R-xy}, \protect\subref{cc-jul-R-xy} July precipitation (\md{}). \protect\subref{mix-Rdiff-xy} July precipitation difference (\md{}) (\cc{}--(\gp{}+\cm{})/2).}
   \label{fig: rean-cam3-Jul-R-xy}
\end{figure}

Figs. \ref{gpcp-jul-R-xy}-\subref*{cc-jul-R-xy} show the precipitation in the month of July of \gp{}, \cm{} and \cc{}. The difference between \cc{} and mean of \gp{} and \cm{} is shown in Fig. \ref{mix-Rdiff-xy}. We have averaged the observational data since the precipitation patterns are very similar. The contrast between \rn{} and model simulations is quite striking. Most noticeable discrepancies in model simulations are (i) significantly reduced precipitation in northern Bay of Bengal, East Asia, western Pacific warm pool (ii) a significant precipitation tongue just south of the equator between 50\de{}-100\de{} and (iii) Spurious precipitation in the Saudi Arabian region which is quite prominent and equal in magnitude to the precipitation peaks in central Arabian Sea and south-western Bay of Bengal. This implies a major realignment in the local heating pattern. This unrealistic precipitation has been discussed by \cite{hurrell-06}.

Thus from table \ref{tab: mix-Umax,R} and Fig. \ref{fig: rean-cam3-Jul-R-xy} it appears that westward shift in the peak of the simulated precipitation is responsible for the TEJ to be centered east Africa.  The strongest impact seems to be from the significantly high anomalous precipitation in the Saudi Arabian region which could play a part in the extreme westward shift of the TEJ.

To get a more quantitative estimate of the change in the precipitation pattern we compute the centroid of the precipitation (\pc{}) in a region that is spatially quite significant and covers the monsoon region. We choose different regions for \rn{} and \cam{} simulations since the precipitation pattern is spatially different. For \rn{} we take into account the west Pacific warm pool and ignore it for model simulations. The region chosen for \rn{} is 60\de{}-130\de{}, 16\ds{}-36\dn{} and for the simulations it is 40\de{}-110\de{}, 16\ds{}-36\dn{} (region demarcated by red boxes in Figs. \ref{gpcp-jul-R-xy}-\subref*{cc-jul-R-xy}). From table \ref{tab: mix-Umax,R} we see that the mean precipitation in simulations and \rn{} are in reasonable agreement although \cc{} experiment overestimated the precipitation by $\sim$15\%. The excess precipitation could be a cause for stronger jet speeds. The centroid of the precipitation is computed using the following equation:

\begin{equation}\label{centroid}
x_c = \frac{\sum_i{P_ix_i}}{\sum_i{P_i}}, y_c =  \frac{\sum_i{P_iy_i}}{\sum_i{P_i}}
\end{equation}

\noindent{where,} \\
$x_c$ and $y_c$ are the zonal and meridional coordinates of the \pc{}, \\
$P_i$ is the precipitation at each grid point, \\
$x_i$ and $y_i$ are the zonal and meridional distances from a fixed coordinate system, in each case the grid point where peak precipitation occurs.

The distances are measured from a coordinate system centered at a point where the precipitation is maximum in the chosen region. There is a clear $\sim$20\degrees{} westward shift in the precipitation centroid in \cc{} with respect to \rn{}. Precipitation pattern of \nglo{} is similar to \cc{} and hence even in this case the shift is $\sim$20\degrees{} westward. The latitudinal differences are not significant. Choosing slightly different averaging regions does not significantly distort this relationship.

\section{Can a realignment in the precipitation patterns cause the simulated TEJ to shift to its correct position?}\label{zm}

The extreme westward shift of the simulated TEJ also leaves open the question whether the heating due to \tp{} is too far away to influence the jet. These two simulations offer the view that latent heating rather than orographic effects is be dominant. A more revealing experiment would be whether shifting the simulated precipitation to the correct location, with and without orography, can cause the jet to also relocate to its correct location. 

The answer to the above question is in the affirmative. We demonstrate this by making changes in the \zm{} deep-convective scheme of \cam{}. \cite{rasch-06} pointed out that CAM-3.0 severely overestimates the fraction of deep convective rainfall over the tropics (\cam{} reproduces CAM-3.0 climate in control configuration). According to them this overestimation was one of the reasons why the simulated precipitation was unrealistic. We have made two kinds of changes:
\begin{enumerate}[(i)]
\item Changing the convective relaxation time scale for deep convection (\td{}) from the default value of 1 hr to 12 hours. This parameter has an important influence on the precipitation pattern. A more detailed study of the influence of \td{} on the mean tropical climate produced by CAM-3.0 is in \cite{saroj-08}. In this set as well we have conducted two experiments -- one with default orography (\cct{}) and one without orography (\nglot{}). No other changes were made.
\item Replacing \zm{} scheme of \cam{} with that of CAM-5.0 which is a part of \cesm{} (Community Earth System Model, version 1.0.2). This automatically removes the possibility of any claim that the community using GCMs face -- that changes in a new model make it difficult to exactly pin down the reason for a change in the simulation. An additional benefit is that we have managed to incorporate some of the most important improvements from a more modern model into our relatively older \cam{}.
\end{enumerate}

The deep convective time scale influences the cumulus scheme in the following way:

\begin{equation}\label{zmeq}
\frac{\partial{}A}{\partial{}t} = -M_bF
\end{equation}

\noindent{where,} \\
$A$: convective available potential energy (CAPE), \\
$F$: CAPE consumption rate per unit cloud updraft base mass flux, \\
$M_b = \frac{A}{\tau{}_dF}$ is cloud base mass flux, \\

The solution to equation \eqref{zmeq} is:
\begin{align*}
A = A_0\exp{}\left(-\frac{t}{\tau{}_d}\right)
\end{align*}

\noindent{where,} \\
$A_0$: initial value of CAPE

Deep convection occurs if CAPE is computed to be above a threshold value which is fixed at 70. We see that a specified adjustment time scale, \td{}, determines how an ascending parcel from the sub-cloud layer acts to destroy the CAPE at an exponential rate. When the relaxation time scale is increased it will reduce the rate at which CAPE is consumed.

\subsection{Changes to \zm{} scheme of \cam{}}\label{zm-cam}

\begin{figure}[htbp]
   \begin{center}
   \subfloat[\cct{}]{\label{cc12td-jul-R,Unorm-xy}\includegraphics[trim = 5mm 20mm 0mm 60mm, clip, scale=0.35]{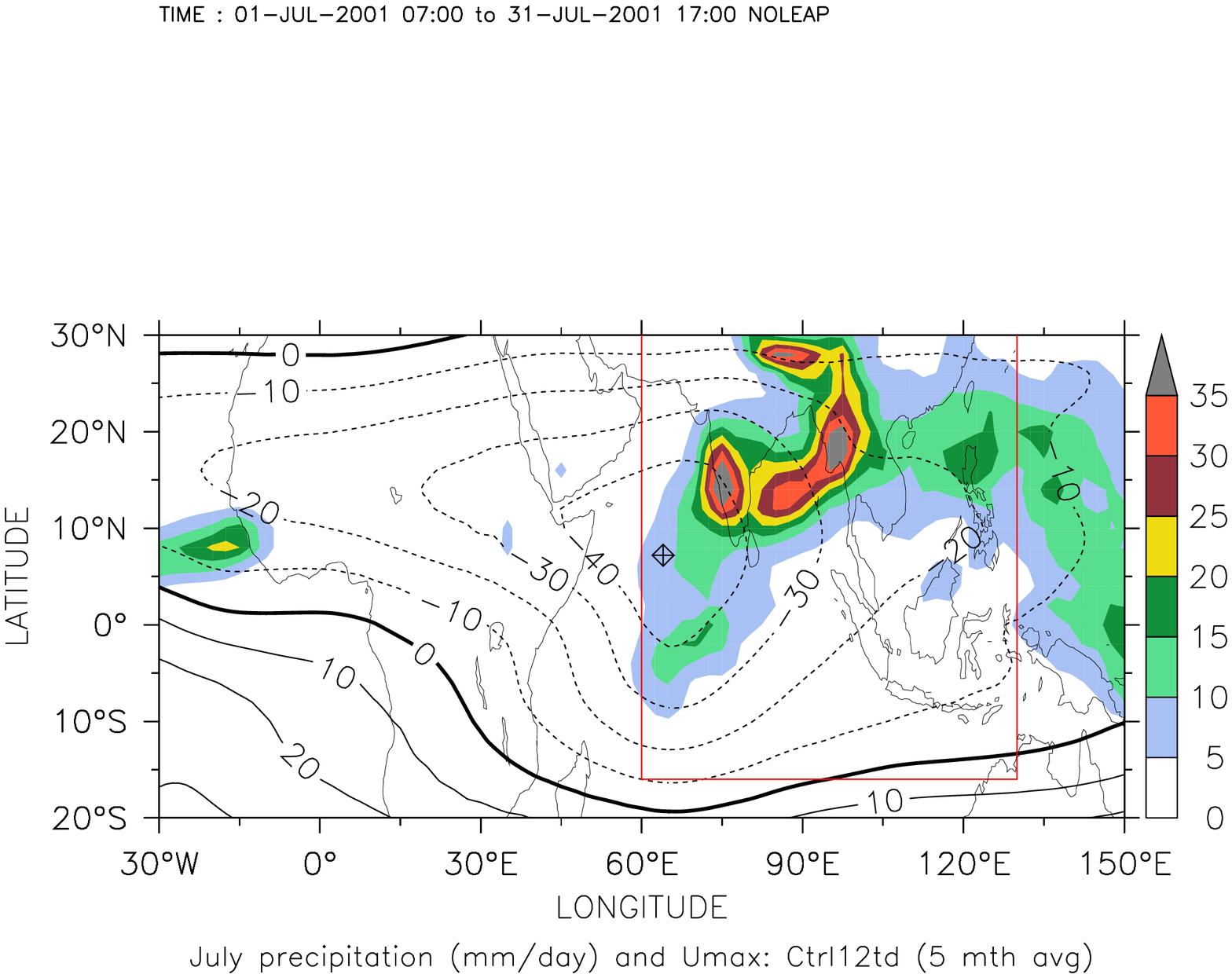}}
   \hspace{5mm}
   \subfloat[\nglot{}]{\label{nglo12td-jul-R,Unorm-xy}\includegraphics[trim = 5mm 20mm 0mm 60mm, clip, scale=0.35]{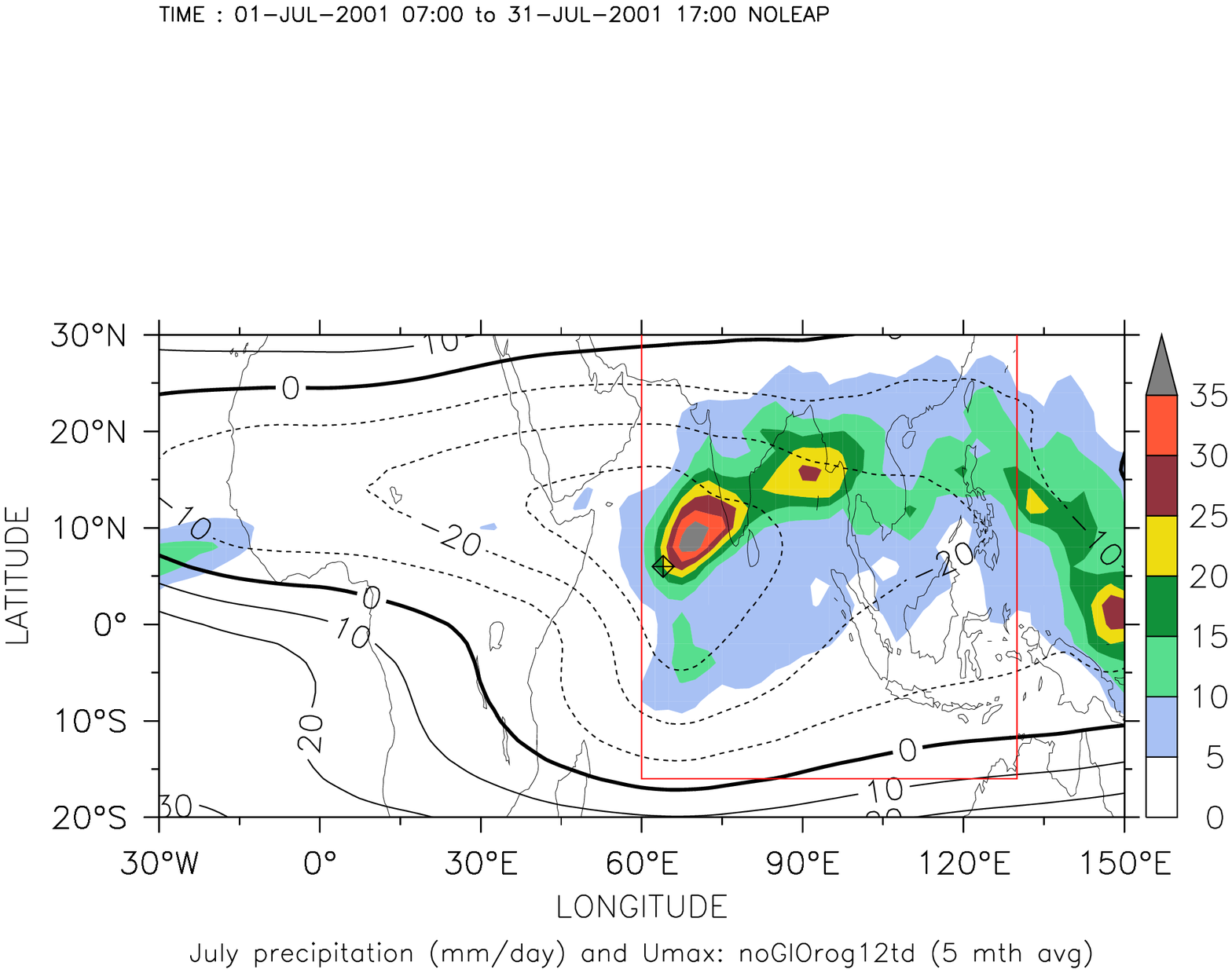}}
   \vskip 5mm
   \subfloat[\ccz{}]{\label{cczm-jul-R,Unorm-xy}\includegraphics[trim = 5mm 20mm 0mm 60mm, clip, scale=0.35]{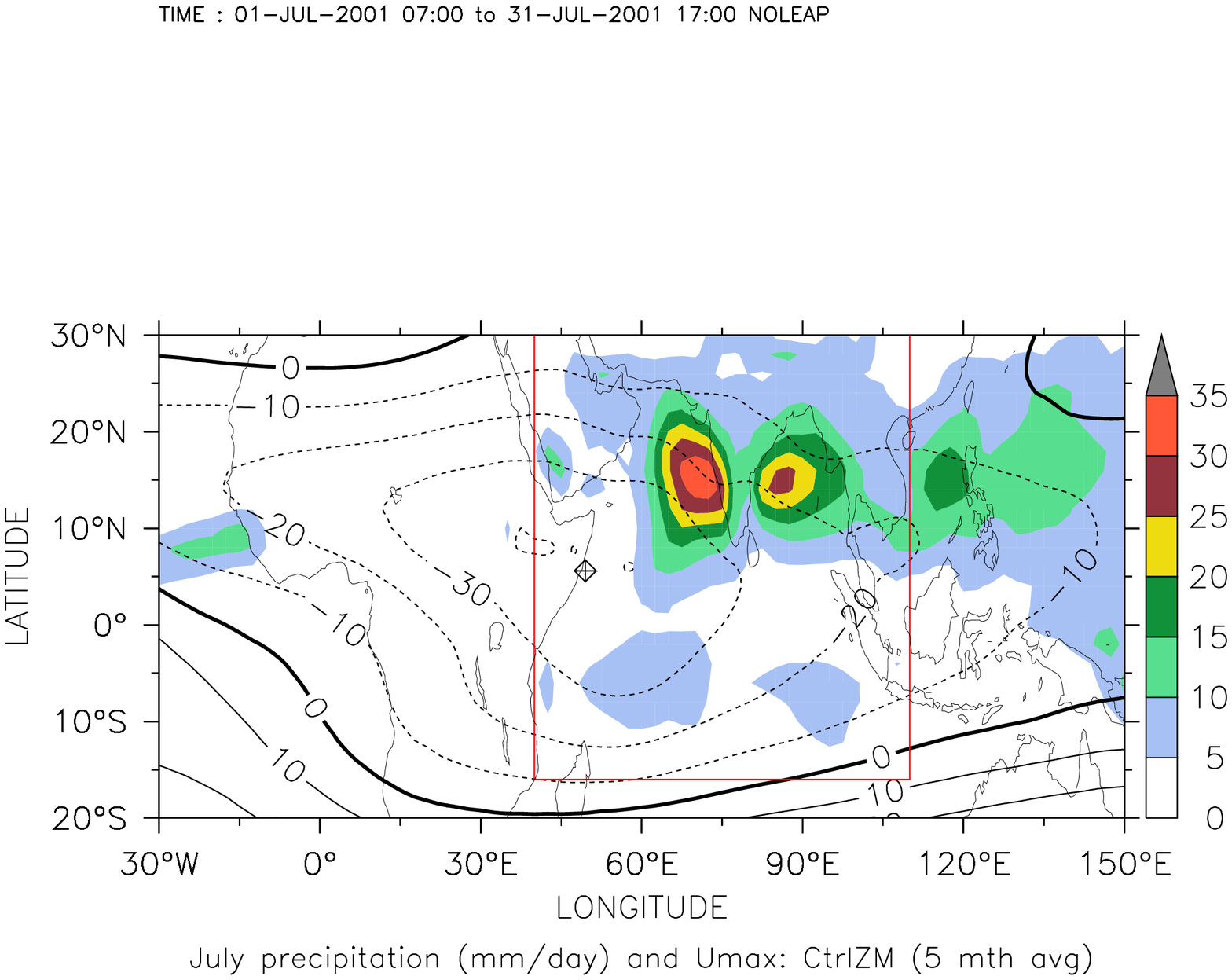}}
   \hspace{5mm}
   \subfloat[\cczt{}]{\label{cczm3td-jul-R,Unorm-xy}\includegraphics[trim = 5mm 20mm 0mm 60mm, clip, scale=0.35]{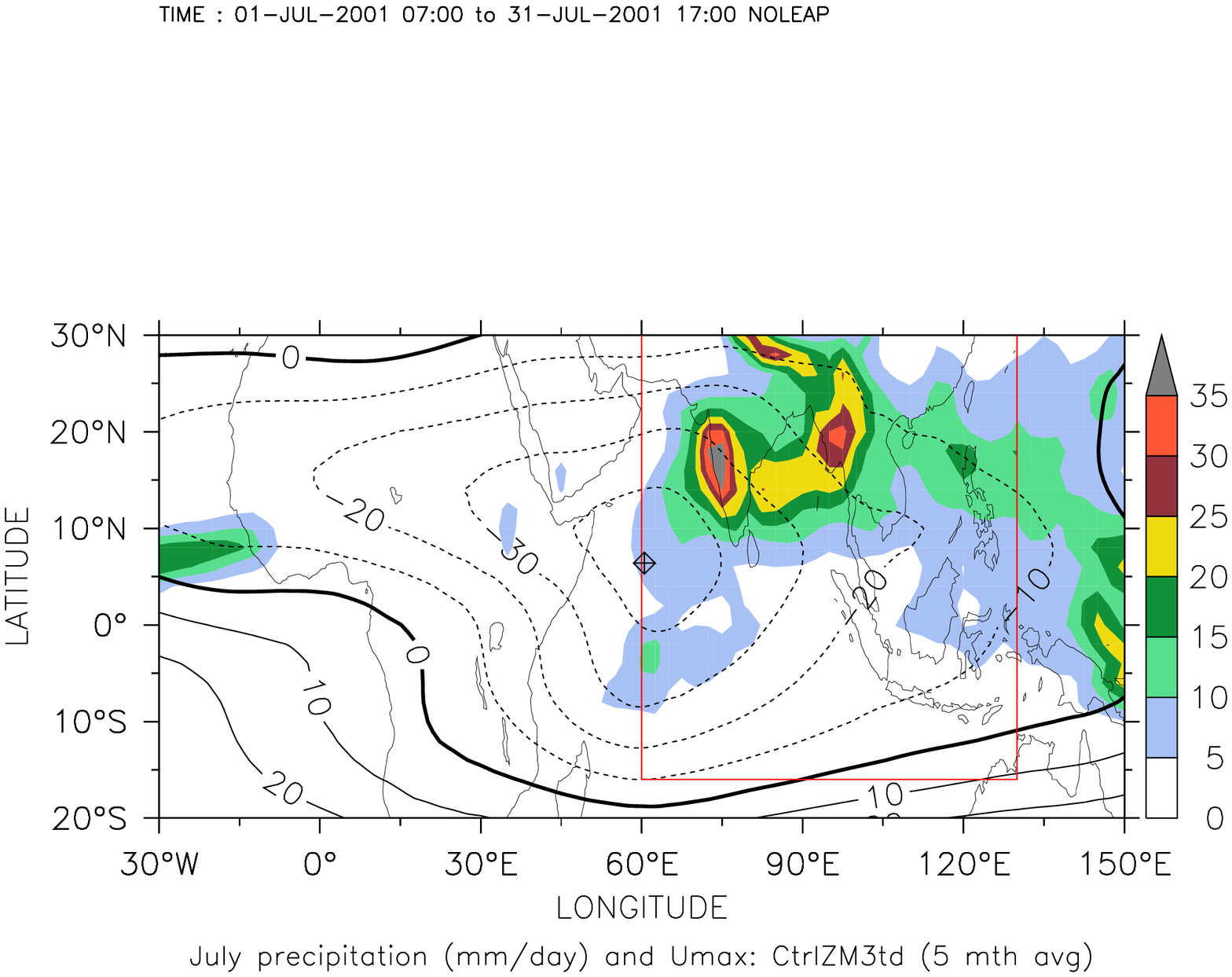}}
   \end{center}
   \caption[July precipitation and zonal wind of \cam{} simulations with modified \zm{} scheme]{July precipitation (\md{}, shaded) and zonal wind of \cam{} simulations with modified \zm{} scheme. \protect\subref{cc12td-jul-R,Unorm-xy}, \protect\subref{nglo12td-jul-R,Unorm-xy}: default \zm{} scheme with \td{} = 12 hours; \protect\subref{cczm-jul-R,Unorm-xy}, \protect\subref{cczm3td-jul-R,Unorm-xy}: \zm{} scheme from \cesm{} with \td{} = 1 hour and 3 hours respectively. Cross-sections are at \um{} pressure level (refer table \ref{tab: mix-Umax,R}), `cross-diamond' shows the location of maximum zonal wind.}
   \label{fig: cam3zm-R,Unorm-xy}
\end{figure}

This corresponds to point (i) above. The precipitation and horizontal zonal wind profiles (at \um{} pressure level) are shown in Figs. \ref{cc12td-jul-R,Unorm-xy},\subref*{nglo12td-jul-R,Unorm-xy}. The precipitation pattern has totally changed. Comparison with \gp{} and \cm{} (Figs. \ref{gpcp-jul-R-xy},\subref*{cmap-jul-R-xy}) shows the spatial pattern to be more realistic. The equatorial precipitation has reduced significantly and that over Western Ghats and Bay of Bengal has increased. Significant precipitation is also observed over the Pacific warm pool. These patterns are observed both in \cct{} and \nglot{} simulations. The only difference is the absence of precipitation to the north of 25\dn{} around 90\de{} in the latter.

\subsubsection{Relative impact of latent heating and orography}

In both simulations, with and without orography, the TEJ is now in the correct location and matches closely with \nc{} (Fig. \ref{nc-Umax-xy}). Table \ref{tab: mix-Umax,R} also shows that the locations of \um{} and precipitation centroid are in good agreement. The lack of difference in spatial structure of the jet between \cct{} and \nglot{} again confirms that orography has a limited role to play in maintaining the jet. Thus it can be inferred that if the spatial pattern of rainfall is similar to observations then the TEJ will be correctly located. In other words latent heating determines the position of the jet. The only difference lies in the magnitude of zonal winds speeds which is $\sim$10\ms{} less in \nglot{} simulation which is likely to be related to the reduction in precipitation. Overall the jet shows the same characteristics of narrow entry and exit region. The meridional wind profile of \cct{} simulation also has the same shape as \nc{} and \cc{} which again shows that the shift in the precipitation pattern has only altered the location of the TEJ. The most significant change in rainfall pattern is, however, the total absence of precipitation in the Saudi Arabian region.

\begin{figure}[htbp]
   \begin{center}
   \subfloat[\nc{}]{\label{nc-z3-xy}\includegraphics[trim = 5mm 20mm 10mm 55mm, clip, scale=0.35]{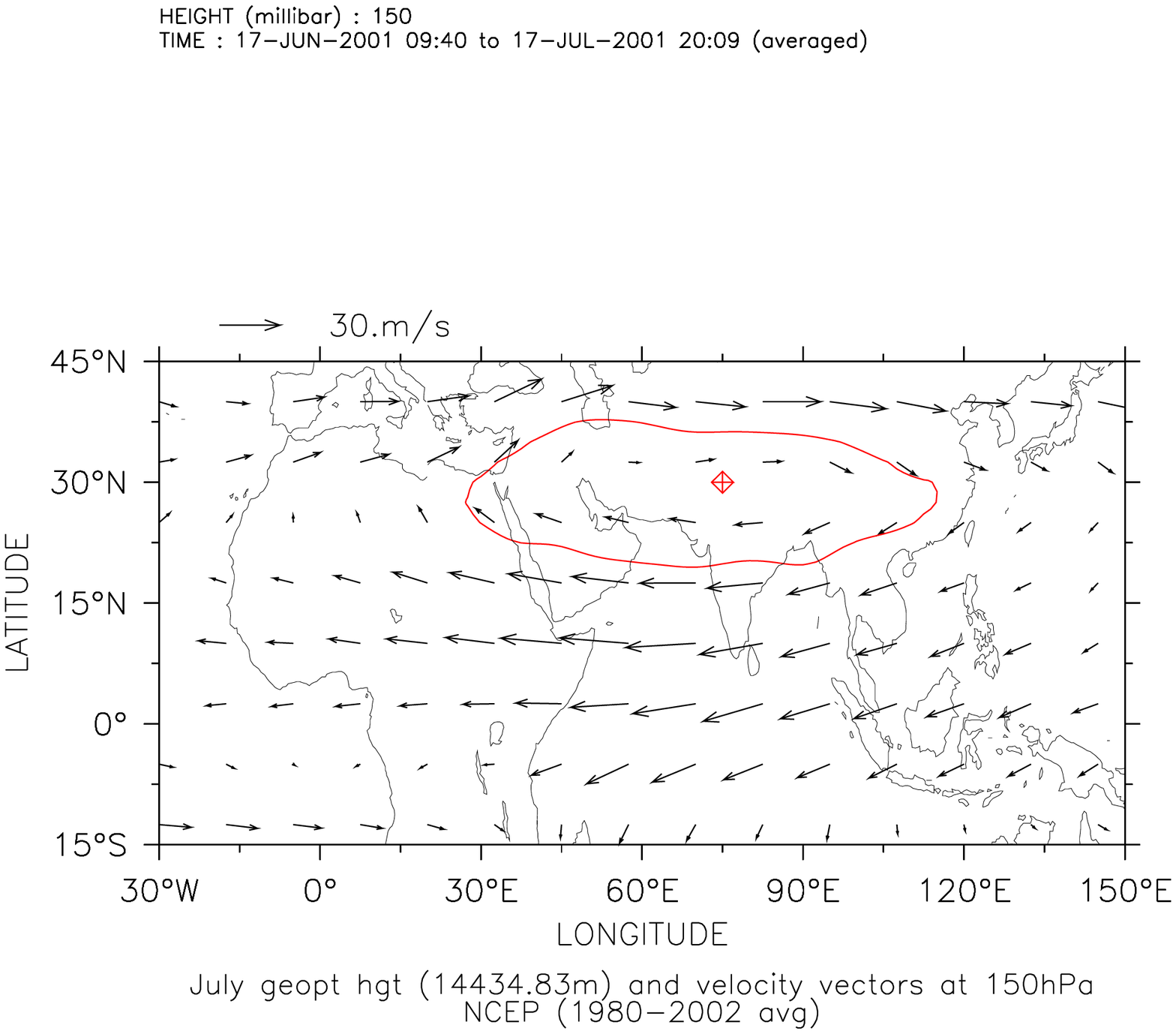}}
   \hspace{5mm}
   \subfloat[\cc{}]{\label{cc-z3-xy}\includegraphics[trim = 5mm 20mm 10mm 55mm, clip, scale=0.35]{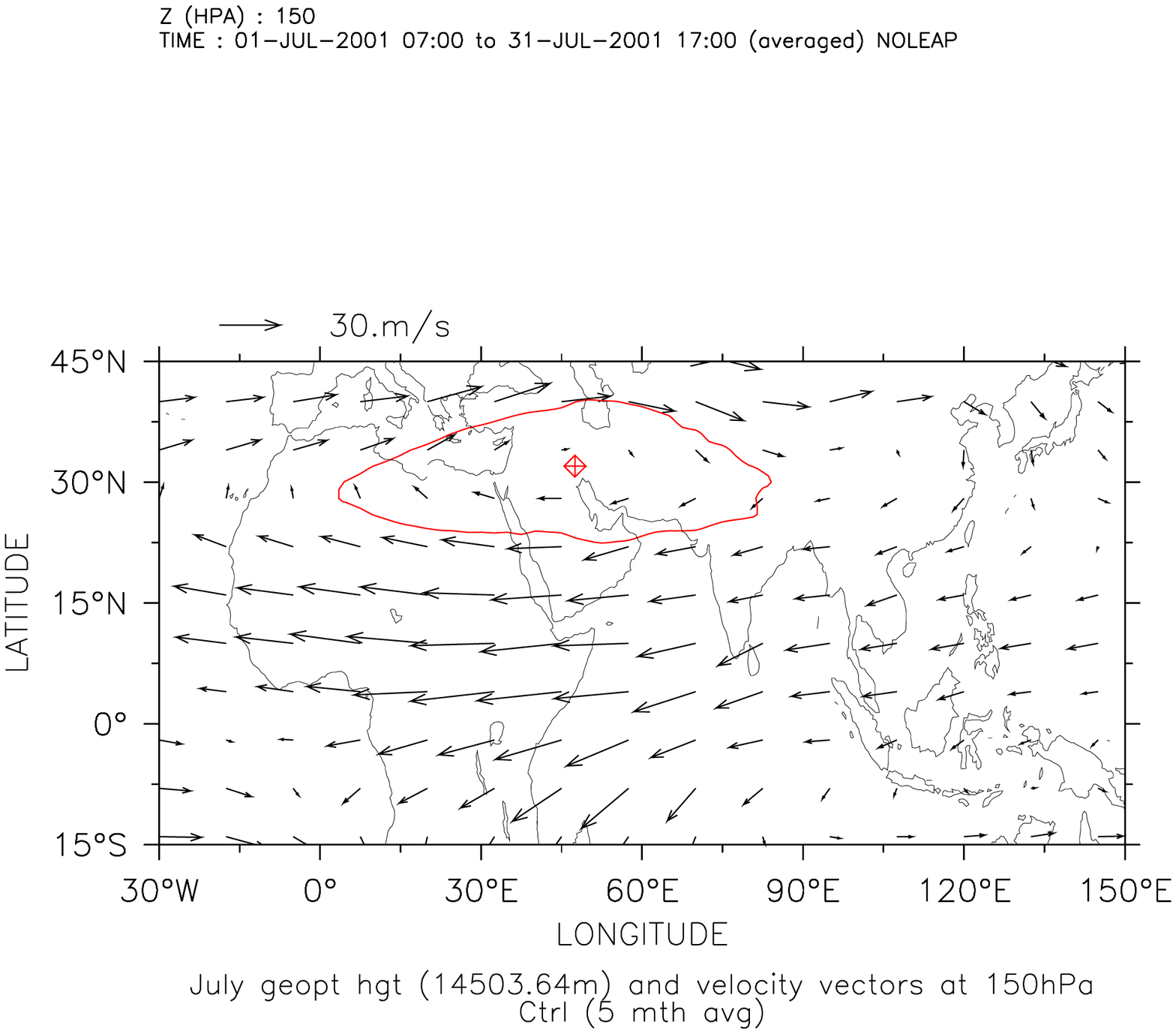}}
   \vskip 5mm
   \subfloat[\cct{}]{\label{cct-z3-xy}\includegraphics[trim = 5mm 20mm 0mm 55mm, clip, scale=0.35]{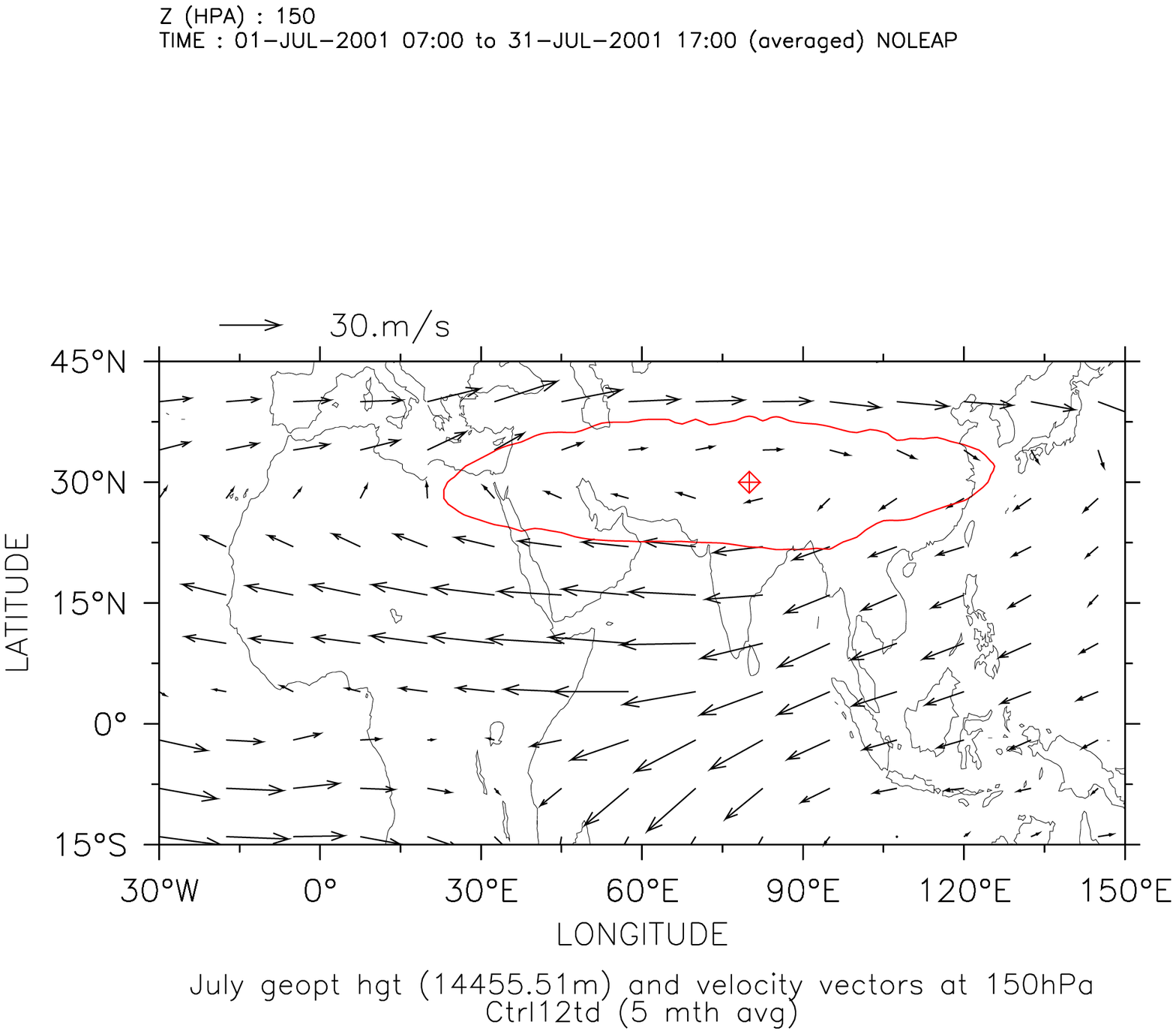}}
   \end{center}
   \caption[July velocity vectors and geopotential height at pressure level where \um{} is attained: \nc{} and \cc{}.]{July velocity vectors and geopotential height ($\times10^{3}$m) at 150 hPa. `Cross-diamond' is location of peak geopotential height. Contour for \nc{}, \cc{} and \cct{} is 14.36, 14.43 and 14.38 respectively which is 99.5\% of individual peak.}
   \label{fig: mix-z3-xy}
\end{figure}

The eastward shift in the heating should also be reflected in the geopotential height. This is clearly seen in Fig. \ref{cct-z3-xy} where geopotential contours at 150hPa are in closer agreement with \nc{} (Fig. \ref{nc-z3-xy}) and distinct from \cc{} (Fig. \ref{cc-z3-xy}). Since \nglot{} simulation is similar to \cct{} we have not shown it.

\subsection{\cam{} simulation with \zm{} scheme of \cesm{}}\label{zm-cesm}

This corresponds to point (ii) above. We have completely replaced \cam{} deep convective scheme with that from \cesm{}. We integrated the model with the initial and boundary conditions of the previous simulations and over the same 5 year time period, we found a significant change in July precipitation both spatially and in magnitude. This we have shown in Fig. \ref{cczm-jul-R,Unorm-xy}. This simulation has been named \ccz{} where `ZM' stands for \zm{}. There has been a major improvement in some aspects. Precipitation in the Saudi Arabian region has reduced in comparison to Ctrl (Fig. \ref{cc-jul-R-xy}). The anomalous precipitation tongue just south of the Equator has also reduced significantly while that in the Pacific warm pool is more realistic (compare with Figs. \ref{gpcp-jul-R-xy},\subref*{cmap-jul-R-xy}).

The zonal wind is displaced $\sim$7\degrees{} eastwards. This eastward shift also hints at the fact that a reduction in heating in Saudi Arabian region causes the jet location to become better. The magnitude and centroid of precipitation (region chosen is same as \cc{} and \nglo{}, 40\de{}-110\de{}, 16\ds{}-36\dn{}) have been calculated as before from Eq. \eqref{centroid}. The meridional wind profile is similar.

The eastward shift in the jet is not particularly dramatic as in \cct{}. On the other hand in order to simulate the location of the TEJ more realistically we had to increase \td{} to 12 hours which is again unrealistic. The next logical step was to change the value of \td{}, in the new convective scheme and check if the desired response is achieved. On changing \td{} from the default value of 1 hour to 3 hours itself the precipitation became similar to \cct{} with the added benefit that the overall magnitude of precipitation also reduced in comparison. This simulation is named \cczt{}. As we can observe from Figs. \ref{cc12td-jul-R,Unorm-xy},\subref*{cczm3td-jul-R,Unorm-xy} there is virtually no change the zonal wind structure and precipitation pattern between \cczt{} and \cct{} simulations. The geopotential contours have also been found to be very similar to \cct{} simulation.

Thus we conclude that the spatial distribution of heating as well as its magnitude is easily the most important factor influencing the location and structure of the \tej{}. Keeping everything else same and only changing that portion of the model which changes the forcing will determine the shape and location of the TEJ. Going with our past experience it can also be concluded that with this new deep convective scheme the jet will not be influenced by orography.

\section{Conclusions}\label{con}

We have studied the July climatological structure of the \tej{} in observations and compared it with the simulations by an AGCM. When we compared the TEJ in \rn{} in July 1998 and 2002 we found that its location had significant zonal shifts. Reduced precipitation in the Indian region in 2002 made the jet have its maxima in the southern Indian peninsula as compared to 1988 when high rainfall in the Indian region resulted in the TEJ having its maxima over the Arabian sea region.

The TEJ in the \cc{} (control) simulation has errors in the spatial location; otherwise the simulated TEJ bears similarities with observations. When we remove orography the spatial location and structure is practically unchanged. This leads us to conclude that the TEJ is not directly influenced by orography. We found that the primary reason for the shift in the simulated TEJ is that location of the precipitation in both \cc{} and \nglo{} is westwards when compared to \rn{}.

We conducted additional experiments to check if the TEJ is primarily influenced by latent heating. Changing the deep-convective relaxation time scale both in \cc{} and \nglo{} simulations confirmed this. In these new simulations the precipitation is more accurately simulated and most importantly anomalous precipitation in Saudi Arabia no longer occurred. The jet followed the shift in precipitation and relocated to the correct climatological position. The absence of orography once again had no impact on the location of the jet. This conclusively proves that the TEJ is independent of orography. Replacing the default deep-convective cumulus scheme with that of \cesm{} also has the same effect on the jet. Both \cct{} and \cczt{} have very similar precipitation patterns and hence the TEJ in both is correctly simulated.

\bibliographystyle{usrdef}
\bibliography{references}

\end{document}